\documentclass[fleqn,usenatbib]{mnras}

\usepackage{newtxtext,newtxmath}

\usepackage[T1]{fontenc}

\DeclareRobustCommand{\VAN}[3]{#2}
\let\VANthebibliography\thebibliography
\def\thebibliography{\DeclareRobustCommand{\VAN}[3]{##3}\VANthebibliography}

\usepackage{graphicx}	
\usepackage{amsmath}	
\usepackage{amssymb}	

\title[Catching unphysical lensed source reconstructions]{Auto-identification of unphysical source reconstructions in strong gravitational lens modelling}

\author[J. Maresca et al.]{
Jacob Maresca$^{1}$\thanks{E-mail: jacob.maresca@nottingham.ac.uk},
Simon Dye$^{1}$,
and Nan Li$^{1,2}$
\\
$^{1}$School of Physics \& Astronomy, University of Nottingham, University Park, Nottingham, NG7 2RD, UK\\
$^{2}$National Astronomical Observatories of China, 20A Datun Road, Chaoyang District, Beijing, China\\
}

\date{Accepted XXX. Received YYY; in original form ZZZ}

\pubyear{2015}

\begin{document}
\label{firstpage}
\pagerange{\pageref{firstpage}--\pageref{lastpage}}
\maketitle

\begin{abstract}
With the advent of next-generation surveys and the expectation of discovering huge numbers of strong gravitational lens systems, much effort is being invested into developing automated procedures for handling the data. The several orders of magnitude increase in the number of strong galaxy-galaxy lens systems is an insurmountable challenge for traditional modelling techniques. Whilst machine learning techniques have dramatically improved the efficiency of lens modelling, parametric modelling of the lens mass profile remains an important tool for dealing with complex lensing systems. In particular, source reconstruction methods are necessary to cope with the irregular structure of high-redshift sources. In this paper, we consider a Convolutional Neural Network (CNN) that analyses the outputs of semi-analytic methods which parametrically model the lens mass and linearly reconstruct the source surface brightness distribution. We show the unphysical source reconstructions that arise as a result of incorrectly initialised lens models can be effectively caught by our CNN. Furthermore, the CNN predictions can be used to automatically re-initialise the parametric lens model, avoiding unphysical source reconstructions. The CNN, trained on reconstructions of lensed Sérsic sources, accurately classifies source reconstructions of the same type with a precision $P > 0.99$ and recall $R > 0.99$. The same CNN, without re-training, achieves $P=0.89$ and $R=0.89$ when classifying source reconstructions of more complex lensed HUDF sources. Using the CNN predictions to re-initialise the lens modelling procedure, we achieve a 69 per cent decrease in the occurrence of unphysical source reconstructions. This combined CNN and parametric modelling approach can greatly improve the automation of lens modelling.
\end{abstract}




\section{Introduction}

Galaxy-galaxy strong gravitational lensing is a unique tool for investigating a wide variety of interesting astrophysical questions. Strong lensing has been used to investigate the nature of dark matter, such as placing lower bounds on neutrino masses in sterile neutrino dark matter models \citep{darkmatter}. Strong lensing has been effective in studying the mass profiles of elliptical galaxies both in the local universe and at cosmological scales \citep{Koopmans_2003, lagattuta}. The lensing of extended sources allows for detailed analysis of galaxy density profiles which can provide insights into the dark matter substructure of galaxies \citep{vegetti_koopmans_2009a, vegetti_koopmans_2009b}. Combining strong lensing measurements with other probes, such as spectroscopy has lead to an increased understanding of the evolution of the mass profile in elliptical galaxies over cosmic time \citep{evolution}. Time delay cosmography, where a variable background source such as a quasar is multiply imaged by a lensing galaxy allows for the inference of key cosmological parameters, such as the Hubble constant \citep{tdcosmo}; \cite{wong}.

In addition to learning about massive elliptical galaxies, strong lensing allows us to probe populations of high redshift source galaxies \citep{richard_2011, dye2018}. Spatially resolved observations of strongly lensed star-forming galaxies enable the study of kinematics on sub-kpc scales \citep{rizzo, jones_2010, swinbank_2009}. High-resolution interferometers such as the Atacama Large Millimetre Array (ALMA) have made it possible to study these sources in exquisite detail \citep{sdp81}. 

There have been several surveys with a focus on lensing, such as the Sloan Lens ACS (SLACS) survey \citep{bolton2006sloan}, the CFHTLS Strong Lensing Legacy Survey (SL2S; \cite{CFHTLS} and the BOSS Emission Line Lens Survey (BELLS; \cite{bells}. To date, the number of strong lensing systems we know of is still relatively small, measuring in the hundreds. This is set to change in the coming years, with two significant surveys coming online. Euclid \citep{euclid}, the European Space Agency's telescope scheduled to launch in 2022 will cover 15,000 $\mathrm{deg}^2$ over 6 years and study the accelerated expansion of the universe out to a redshift of $z = 2$. Additionally, the Vera Rubin Legacy Survey of Space and Time \citep[LSST;][]{lsst}, also focused on the study of dark energy and dark matter, will commence science operations in 2023. LSST will cover 18,000 $\mathrm{deg}^2$ over ten years in six different filters ($u, g, r, i, z, y$). It is expected that these surveys will discover many tens of thousands of lensing systems; 120,000 and 170,000 lenses for LSST and Euclid respectively \citep{collet}. For this reason, the development of fast, automated, and accurate pipelines for finding and modelling strong lenses is of great importance.

Typical methods for finding strong gravitational lenses are based upon visual inspection of candidate images that have been selected using properties such as morphology, colour and luminosity \citep{pawase, sygnet}. Searches for high-redshift spectral lines present in lower redshift galaxies have been used to find strong gravitational lenses, such as in the SLACS survey. Techniques designed to identify arc-like structures and rings in images have been developed and applied to surveys with some success \citep{seidel, Gavazzi_2014}. Approaches based on the quality of fit to the data achieved by lens modelling have been developed \citep{Marshall_2009, yattalens}, although the speed and flexibility of such approaches is a challenge for dealing with large amounts of data. Another approach to this problem utilises supervised machine learning algorithms, such as artificial neural networks and Gaussian mixture models \citep{bom, ostrovski}. Recently, there has been interest in developing unsupervised machine learning algorithms to tackle the challenge of lens finding \citep{sunny}.

Finding strong gravitational lenses is only one aspect of the challenge; they must also be modelled. When dealing with the lensing of an extended source, we wish to reconstruct both the source's intrinsic brightness distribution as well as model the mass distribution of the lens galaxy. One such method is that of semi-linear inversion \citep{SLI}, a technique that reconstructs the pixelised source in a linear step for a given lens model. This technique has been placed within a Bayesian framework for optimising the model evidence \citep{suyu} and more recent implementations reconstruct the source on an irregular grid of pixels that can adapt to the lens magnification or the source surface-brightness \citep{adaptive_sli, Nightingale2018}. Another method for reconstructing the intrinsic source makes use of the family of polynomials known as shapelets \citep{shapelets_birrer}. An analytical reconstruction of the source can be formed using a small subset of these polynomials, leading to a reduced number of source parameters \citep{shapelet_bayes}. Convolutional Neural Networks (CNNs) have been used to reliably and automatically recover the mass-model parameters of galaxy-galaxy strong lenses in orders of magnitude less time than traditional parametric techniques \citep{hezaveh, pearson}. Furthermore, advancements have been made in the application of neural networks for reconstructing the background source of a strongly lensed system \citep{morningstar}.

Techniques that model the lens mass with a parametric density profile remain a necessary and indispensable tool. There are significant difficulties involved in creating unbiased and sufficiently varied training sets for CNNs to learn from. This is a particular problem in the case of lensed high-redshift sources, where the source light is likely to be highly irregular. In addition, contamination of a lens data-set by objects such as galaxy mergers and ring galaxies poses a problem for CNN based methods. In these circumstances, a CNN will produce a set of lens model parameters without any indication of failure, whilst parametric modelling techniques will fail to fit the data since they operate within the context physically-motivated density profiles and are bound by the multiple imaging constraints of a real lens. Typically, it has been necessary to rely upon parametric techniques to obtain a robust measure of uncertainties on the lens parameters. Recently, however, methods for obtaining the uncertainties on CNN predicted parameters have been developed \citep{levasseur, Park2020}.

A particular issue for methods based on the pixelised source reconstruction is the existence of unphysical solutions (see Section \ref{sec:problem} for details). Such solutions are perfectly valid, providing excellent fits to the data and can be challenging for sampling algorithms to avoid. An unsupervised modelling run can spend large amounts of time exploring the parameter space around these solutions and never converge towards the true parameter values. These solutions can be avoided with careful tuning of the model parameters, but this represents a significant investment of time for each system being modelled. For this reason, we have developed a CNN based approach to recognise these unwanted solutions and a simple prescription for updating the priors in our model to aid convergence towards the true solution. In this manner, we can iteratively improve our lens model by identifying and avoiding reconstructions that correspond to under and over-magnified solutions.

The paper is organised as follows: Section \ref{sec:problem} describes the occurrence of unphysical solutions in the modelling process and their properties. Section \ref{sec:method} discusses the methodology for simulating the required images of strongly lensed galaxies, and the processes involved to create source reconstructions from these images. Additionally, an overview of the CNN architecture is provided with details on how the network was trained and the manner in which the CNN was used in conjunction with our modelling process. The results of applying this technique to our testing set of data are presented in Section \ref{sec:results}. Finally, the results in this work are discussed along with our conclusions in Section \ref{sec:conclusions}. Throughout this paper we assume a flat $\Lambda\mathrm{CDM}$ cosmology using the 2015 Planck results \citep{planck2015}, with Hubble parameter $h=0.677$ and matter density parameter $\Omega_m = 0.307$.

\section{Erroneous Solutions and Their Inversions}
\label{sec:problem}

One of the key motivations for using the Semi-linear inversion method is the reduced computational complexity of the lens modelling process. Using analytic profiles to model the complex source light of a lensed galaxy can require exploring a highly multi-dimensional parameter space. Not only does this increase the likelihood of inferring a solution corresponding to a local maximum in evidence, but can also lead to biasing of the lens model due to constraints on the light profiles. Semi-linear inversion allows us to reconstruct the source light distribution in a linear step and since this distribution is pixelised, it is not constrained by an analytic profile. It does however introduce a new set of problems for the modelling process, namely under and over-magnified solutions.

\begin{figure*}
    \includegraphics[width=17cm]{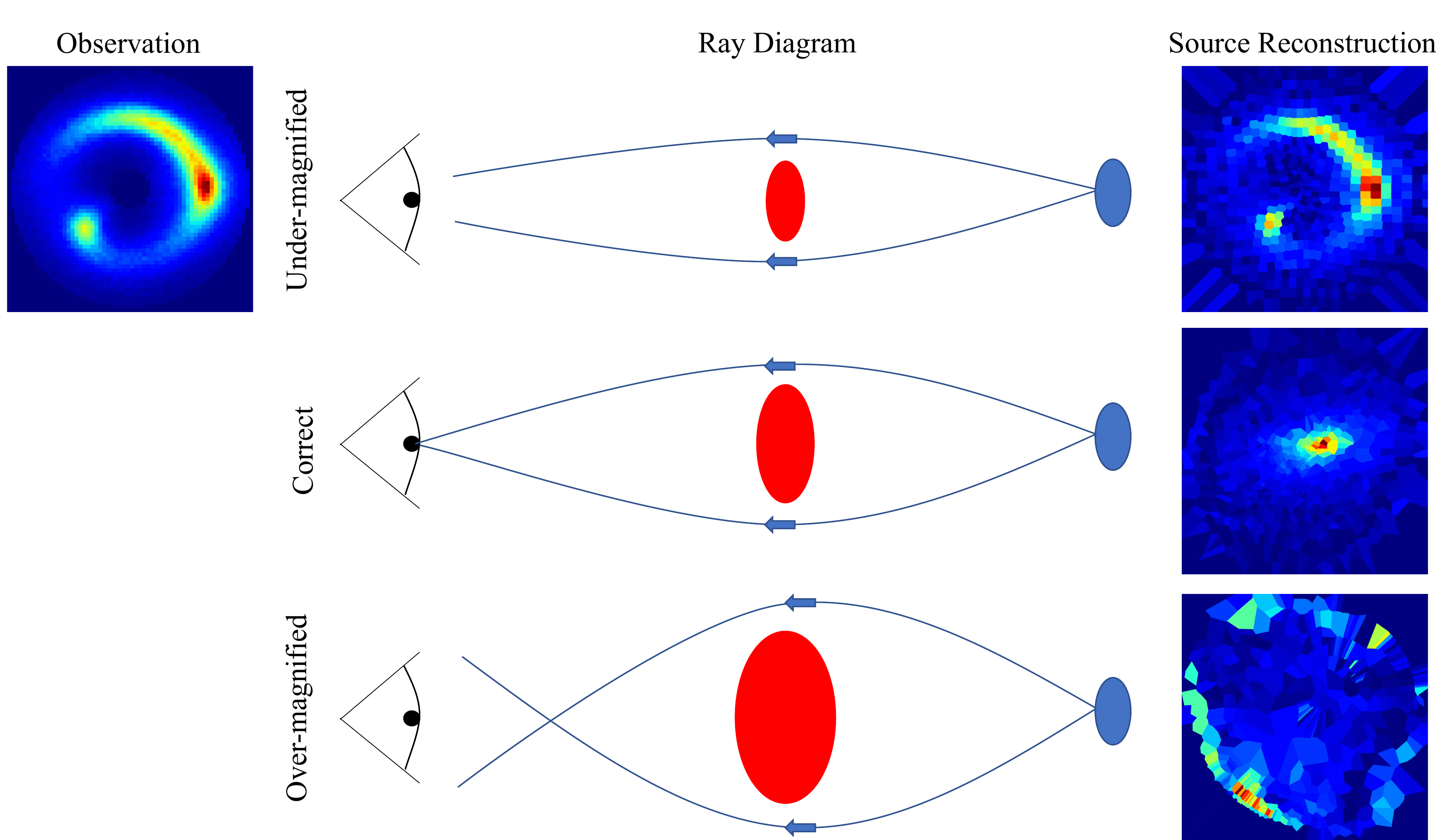}
    \caption{From left to right: Original observation, observer, lens galaxy, source galaxy, reconstructed source. From top to bottom: under-magnified solution, correct solution, over-magnified solution. Note the similarities in structure between the under-magnified solution and the original observation. Both images have similar morphology, but different angular extents. It is more challenging to see the similarities between the over-magnified solution and the original observation, but it is clear that the extended arc in the north of the observation is being reproduced in the south of the reconstruction, illustrating the parity inverted nature of the solution.}
    \label{fig:ray_diagram}
\end{figure*}

These so-called under-magnified and over-magnified solutions can be understood in terms of the inferred amount of mass in the model lens galaxy. Here, we use the Einstein radius as a proxy for the mass in a galaxy. Ideally, the modelling process will converge upon the true value of the Einstein radius, along with the other model parameters, and the reconstructed source will reproduce the unlensed features of the source galaxy. If however, the modelling process converges upon a solution with too small an Einstein radius, the resultant deflection angles will also be under-estimated. This leads to the formation of an under-magnified image of the observation itself. Similarly, a model with too large an Einstein radius will over-estimate the deflection due to the lens. This will lead to an over-magnified, but this time, parity inverted image of the source. Fig.~\ref{fig:ray_diagram} illustrates this point with stylised ray diagrams for each class of source reconstruction we are considering.

\begin{figure*}
    \includegraphics[width=17cm]{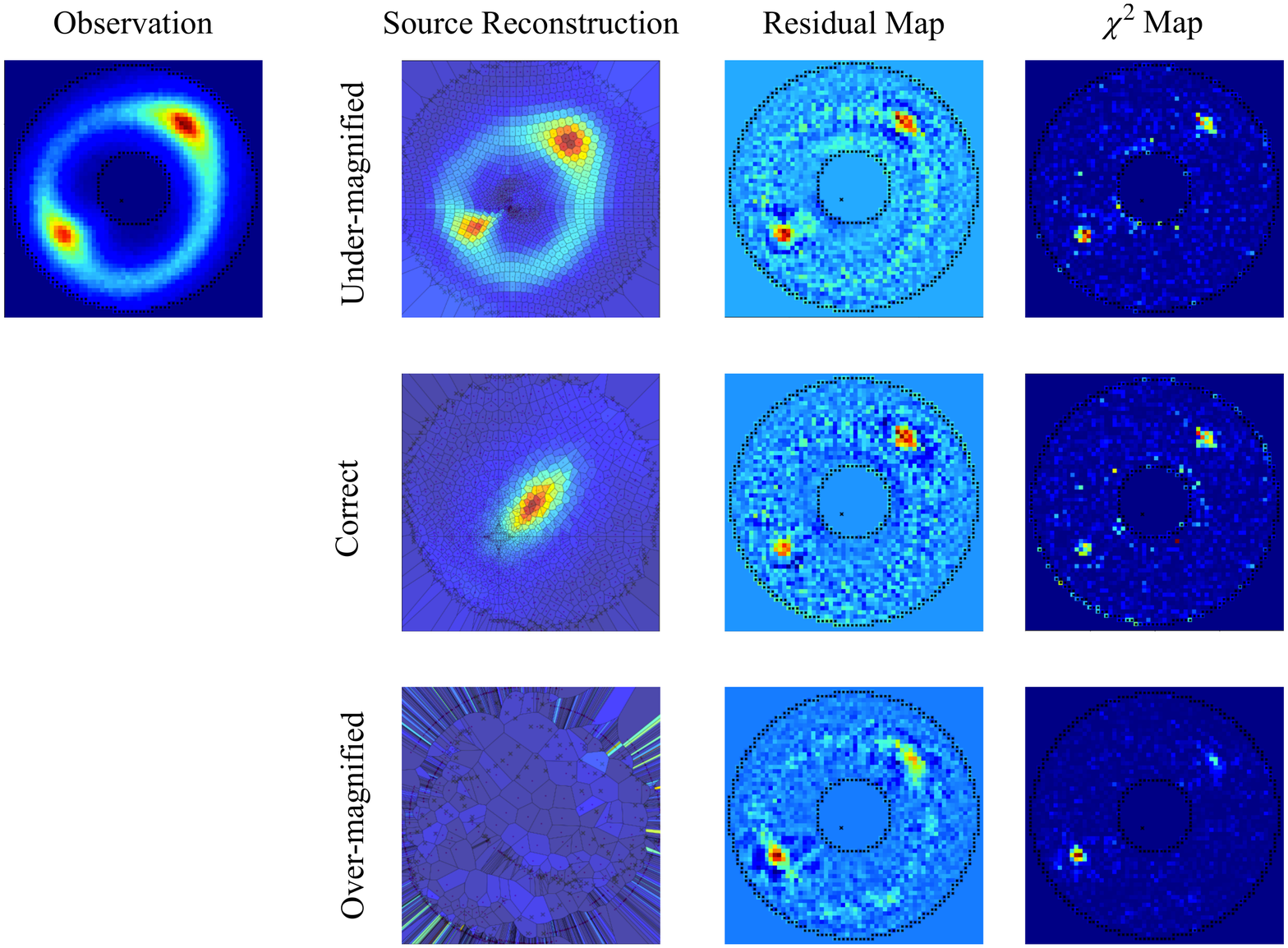}
    \caption{Column 1 contains a simulated observation of a lensed galaxy. Column 2 contains the source reconstructions for three classes of solution (under-magnified: row 1, correct: row 2, over-magnified: row 3). Column 3 contains the residual maps (model image - observation) for each class of solution. Column 4 contains the chi-squared maps (the squared significance of the residuals) for each class of source reconstruction.}
    \label{fig:recon_grid}
\end{figure*}

Whilst these erroneous source reconstructions are obviously not the physical solution we are looking for, they exist nevertheless and can provide excellent fits to the data, thus posing a challenge for sampling algorithms to avoid them. Fig.~\ref{fig:recon_grid} shows an example of another set of source reconstructions for a simulated observation. Here, we also show the residual and chi-squared maps for each reconstruction, showing the quality of the fit to the data. The Bayesian evidence is comparable for the under-magnified solution and the correct solution, whilst it is significantly lower for the over-magnified case. Generally speaking, we find that the under-magnified solution is much more probable to occur than the over-magnified one. This is likely due to the regularisation employed in the semi-linear inversion process. Regularisation serves to penalise overly complex solutions, which is certainly a characteristic of the over-magnified solution. In addition to regularisation reducing the likelihood of this solution, it can often be excluded by sufficiently accurate masking of the lens system. Provided the mask used when modelling the system does not extend considerably farther than the image separation, it can be used to set the upper-bound on the Einstein radius prior.

It is usually clear to the experienced modeller when something has gone wrong and an erroneous source reconstruction has been produced. It is not however so easy to discriminate between these solutions programmatically. Multiple techniques can be employed to avoid these solutions; careful tuning of the prior distributions on the lens model parameters is effective but time-consuming. For this reason, it is not a suitable method for dealing with the large numbers of lensed galaxies we expect to encounter in the coming years. Another possibility, which has the benefit of being automatic, is to create a pipeline of models that first fits an analytic light profile to the source galaxy and then uses the results of this fit to initialise the priors in the inversion process \citep{Nightingale2018}. By requiring a compact source in the initial phase of modelling, the aim is to infer a lens model sufficiently accurately to effectively rule out regions of parameter space that would correspond to under or over-magnified solutions. This lens model, along with new priors on its parameters is then used in the inversion process to refine the lens model and more accurately fit the source galaxy's light. The complex morphology of high-redshift sources poses a challenge for fitting the data with an analytic light profile, which can lead to a poorly constrained or entirely wrong lens model. If the inferred lens parameters used to initialise the model in the inversion process are of poor quality, then the modelling can once again fail at this step. Even if a fit with an analytic source profile provides a reasonable initialisation for the inversion process, it is challenging to constrain the width of the subsequent prior distributions such that erroneous source reconstructions are ruled out but feasible lens models that fit the more complex source are not. Our approach to this challenge is to use a CNN that can accurately classify source reconstructions as successful or under/over-magnified. In this way, we completely remove the need to assume an analytic light profile for the source since we can throw away unwanted solutions in the inversion process that do not correspond to a compact reconstructed source. Furthermore, we have developed a simple method for updating the model to move away from these unwanted solutions towards the correct parameters. This technique requires no human intervention and the CNN classification step is extremely fast (<1s).

\section{Methodology}
\label{sec:method}

The CNN described in this work requires training data consisting of labelled source reconstructions and residual images. To produce this data, it was first necessary to create a large number of simulated strong gravitational lens images. We used the lens modelling software {\tt PyAutoLens} \footnote{https://github.com/Jammy2211/PyAutoLens} \citep{pyautolens, adaptive_sli, Nightingale2018} to produce our simulated images and to perform the source reconstruction. Multinest \citep{multinest} was used for the exploration of parameter space where a full analysis of the data was carried out. The modelling process produces the residual images between the simulated observations and the reconstructed model image that we need for training the CNN. In Section \ref{subsec:sim} we describe our procedures for generating the simulated strongly lensed images. Section \ref{subsec:recon} details our method for generating the source reconstructions and residual images required for training our neural network. We then describe the CNN architecture used in this work in Section \ref{subsec:cnn}. The process used to update the prior distributions on the model, based on the CNN predictions is then detailed in Section \ref{subsec:combo}.

\subsection{Lensing Simulations}
\label{subsec:sim}

\begin{figure}
    \includegraphics[width=\columnwidth]{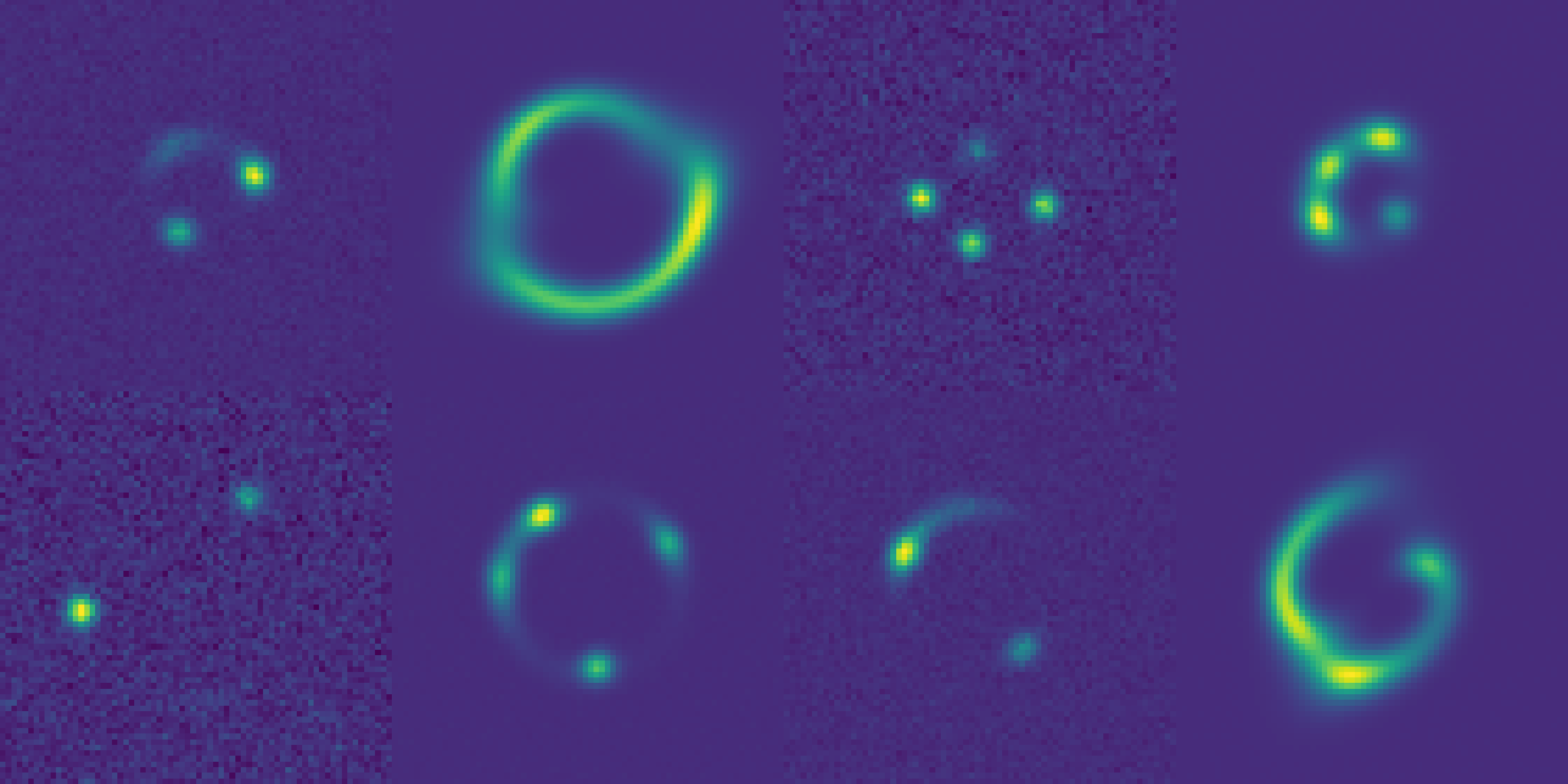}
    \caption{A selection of simulated images produced for this work, used for creating pixelised source reconstructions to train a CNN. All images have a pixel scale of 0.1 arcsec pixel$^{-1}$ and each image's color scale has been normalised to the peak signal of the image.}
    \label{fig:lens_examples}
\end{figure}

In this work, we have assumed that all of the foreground deflectors are early-type galaxies, and so we have adopted the Singular Isothermal Ellipsoid (SIE) mass profile \citep{keeton2001catalog}. For the light profile of the background lensed galaxies, we have opted to use the Sérsic profile since it can represent a wide variety of galaxy morphologies.

The data sets generated for this work were simulated to have distributions of parameters similar to those observed in the Sloan Lens ACS (SLACS) survey \citep{bolton2006sloan}. The Einstein radius $\theta_E$ and axis ratio $q$ of our lensing galaxies were drawn from distributions fitted to the measurements of 131 strongly lensed galaxies observed in the SLACS survey \citep{bolton2008sloan}, whilst the orientation $\phi$ was allowed to vary uniformly over the full range. The Einstein radii of our lenses were drawn from a normal distribution with mean $\mu = 1.16$ and a standard deviation $\sigma = 0.42$. The axis ratios of our SIE profiles were randomly sampled from a normal distribution with mean $\mu = 0.80$ and standard deviation $\sigma=0.16$, in close agreement with empirical studies \citep{koopmans2006}. In all cases, the centroid of the lens was placed in the centre of the image. In this work, we did not include light from the lens galaxies in the simulations.

As with the lenses, the parameters describing our source galaxy sérsic profiles were randomly sampled from fitted distributions. In this case, we used the inferred Sérsic parameters from the parametric source reconstructions of a subset of the SLACS lenses \citep{newtonapj}. The Sérsic indices $n$, of our sources, were randomly drawn from an exponentially modified Gaussian distribution with scale parameter $\lambda=0.723$, mean $\mu=0.71$ and standard deviation $\sigma=0.97$. The effective radii $r_{\mathrm{eff}}$ of our sources were randomly sampled from an exponential distribution with scale parameter $\lambda=6.64$. We allowed the axis ratio of the sources $q_{s}$ to vary uniformly over the range [0.3, 1]. The overall intensity normalisation $I$ of the sources was drawn from a uniform distribution $I \sim U[10, 20]$ electrons $\mathrm{s}^{-1}$, allowing for a wide variety of signal to noise ratios in our training data. The centroid of each source was uniformly distributed in the source plane, with the requirement that it lay inside the Einstein radius of the lens (i.e that there are multiple images).

In the production of our simulated images, we opted to use the pixel scale of the VIS instrument for Euclid (0.1 arcsec $\mathrm{pixel}^{-1}$) and the characteristic exposure time of 565 seconds \citep{cropper}. The lensed image was then convolved with a Gaussian point spread function with a full-width at half-maximum of 0.17 arcseconds. A background sky of 1 electron $s^{-1}$ and Poisson noise due to the background sky and source light photon counts were added to the images, thus completing the simulation procedure. Some examples of our simulated images are shown in Fig.~\ref{fig:lens_examples}.
\subsection{Training Data}
\label{subsec:recon}

The CNN was not trained directly on the simulated images, but rather the pixelised source reconstructions and residual images obtained from the modelling process. Before the modelling began, each simulated image needed to be masked to ensure that only the area of interest was reconstructed in the source plane and to reduce the computational load. Due to the large number of simulated images, an automated masking scheme was used. Firstly, the images were thresholded using the minimum cross-entropy approach \citep{threshold}. Then, the centroid of this thresholded image was found through calculating its moments. A circular annular mask, centred on the centroid of the image was then fitted to the thresholded pixels. For the inner radius of the annulus, the largest radius circle that did not contain any unmasked pixels was found, and 90 per cent of this value was used. Similarly, for the outer radius, the smallest circle containing all of the unmasked pixels was computed, and 110 per cent of this value was used. These adjusted values for the inner and outer radii of the mask were used to minimise the chances of masking out faint emission from the source.

These masked images were then modelled using {\tt PyAutoLens} to produce the pixelised source reconstructions and residual images that we need for training our CNN. In all cases, we adopted the SIE mass profile to model the lens galaxy. We reconstructed the background source on a pixelised grid that adapts to the magnification of the system. For each simulated lensed image, we created three source reconstructions and three residual images, corresponding to the under-magnified, over-magnified, and correct solutions. This resulted in approximately 300,000 images to be used as training data for our network. To deal with such a large computational task, it was necessary to employ some approximate methods in the source reconstruction/lens modelling process.

For 250 of our simulated images, we performed a full analysis of the data, optimising the lens model and source parameters in the inversion process. In each case, the analysis had to be repeated three times, to produce the under-magnified, correct and over-magnified source reconstructions. When modelling each of these systems we allowed the mass-model parameters to vary uniformly over the full range of parameter space with the exception of the Einstein radius. To produce an under-magnified source reconstruction, we set a uniform prior distribution on the Einstein radius with an upper limit of 0.9 times the true value for the system, thus forcing {\tt PyAutoLens} to find the under-magnified solution. To produce source reconstructions corresponding to the correct model, we allowed the Einstein radius to vary over a small range centred on its true value, guaranteeing that a sensible source reconstruction is produced. Finally, to produce over-magnified source reconstructions, we allowed the Einstein radius to vary over a range of 1.1 times the true value up to 3 times this value, again forcing {\tt PyAutoLens} to find the over-magnified solution. In this manner, we built up an understanding of the properties of each class of source reconstruction.

\begin{figure}
    \includegraphics[width=\columnwidth]{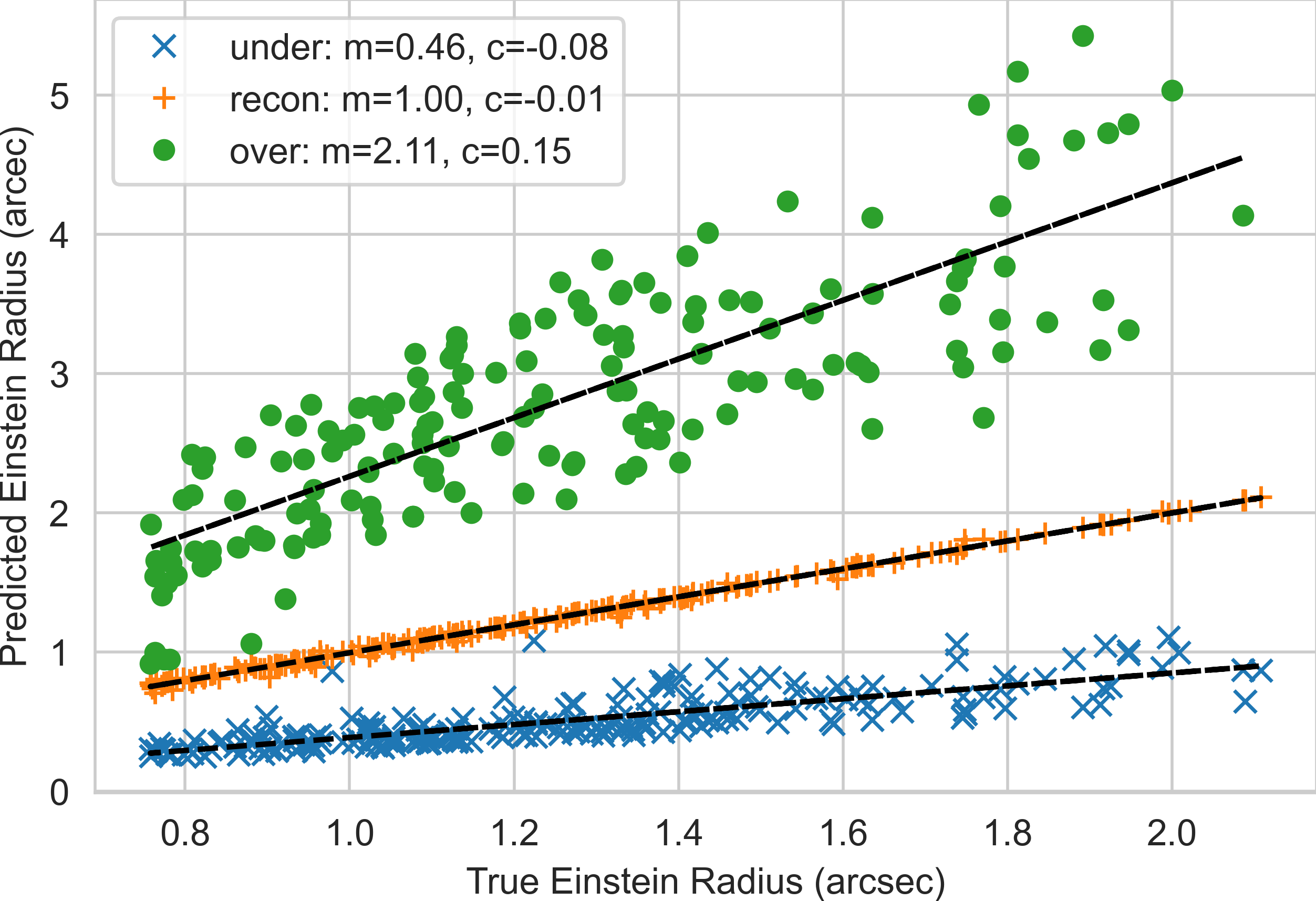}
    \caption{The relationship between the true value of $\theta_E$ and the predicted value $\hat \theta_E$ follows a predictable pattern for each class of source reconstruction. In each case, the coefficients of a linear fit are shown; These are used to construct the transformation of the prior distribution on $\theta_E$ to converge upon the correct solution.}
    \label{fig:pred_true}
\end{figure}

\begin{figure}
    \includegraphics[width=\columnwidth]{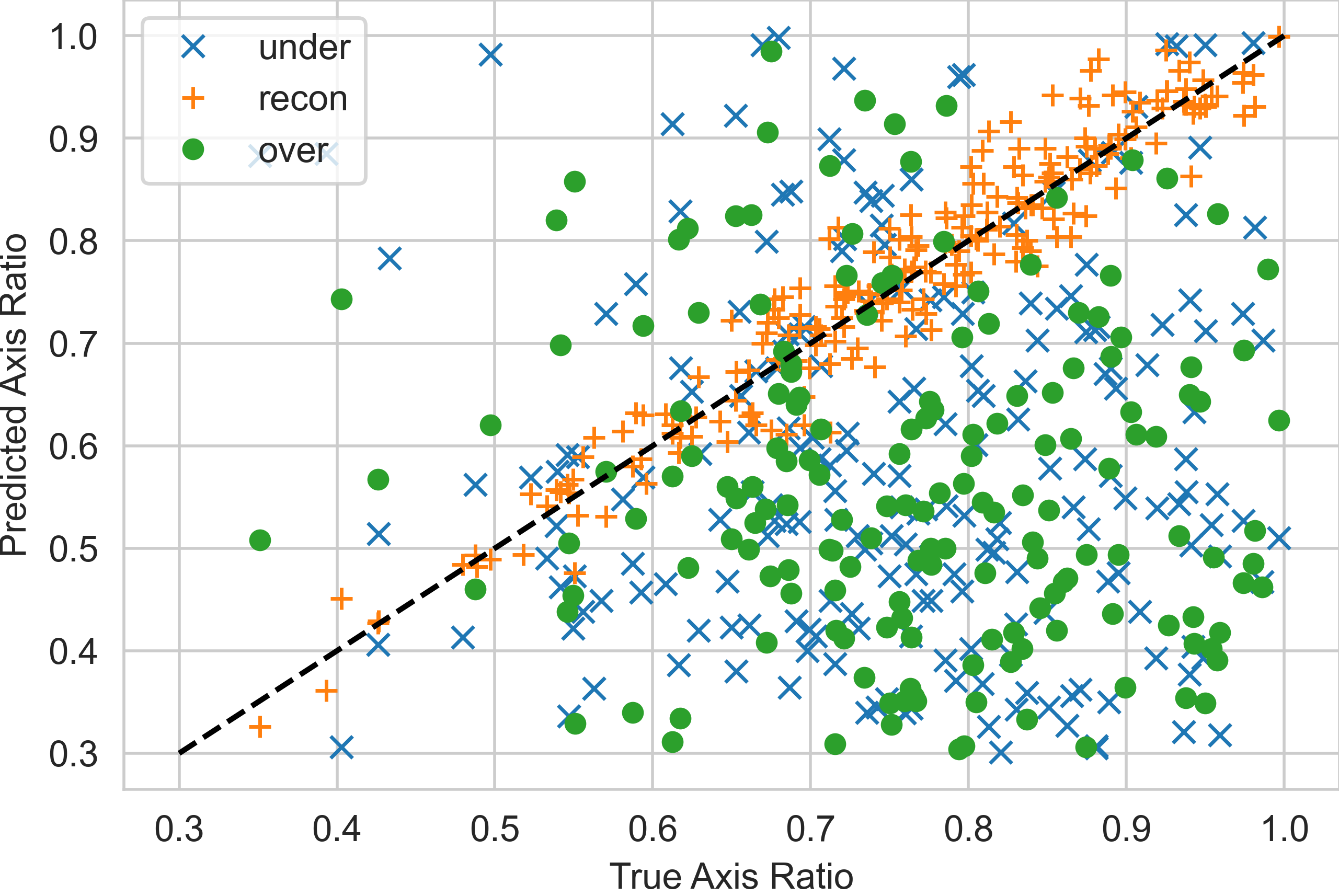}
    \caption{The relationship between the true axis ratio of the lens $q$ and the inferred value $\hat q$ for each class of source reconstruction. For successful source reconstructions, $q$ and $\hat q$ are directly proportional to one another, but in the case of under/over-magnified solutions there is no obvious trend aside from a tendency for $\hat q$ to under-estimate $q$.}
    \label{fig:pred_true_q}
\end{figure}

\begin{figure}
    \includegraphics[width=\columnwidth]{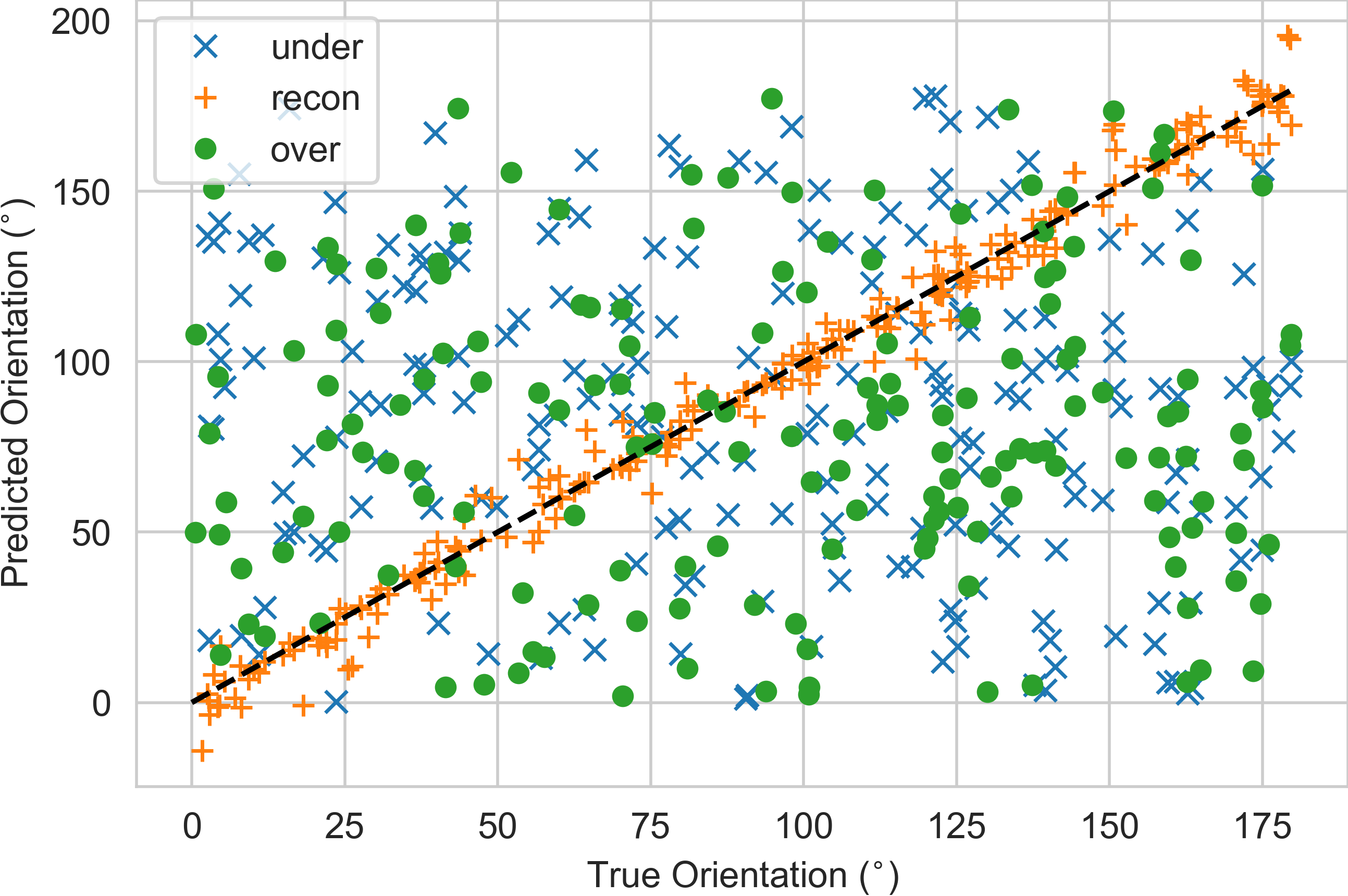}
    \caption{Successful source reconstructions accurately recover the lens orientation, $\phi$, whilst under/over-magnified solutions infer a value $\hat \phi$ that appears to have no relationship to $\phi$.}
    \label{fig:pred_true_phi}
\end{figure}

In these tests, we observed that the mean fractional error in Einstein radius when producing an under-magnified source reconstruction is $\hat f_{\theta_E} \approx -0.5$. As expected, we observed no significant bias in the Einstein radius, or any of the other parameters, when using a model with priors accurately centred on the true parameter values. The mean fractional error in Einstein radius when producing over-magnified reconstructions was $\hat f_{\theta_E} \approx 2$. A scatter plot of the true value of Einstein radius versus the inferred value for each class of source reconstruction is shown in Fig.~\ref{fig:pred_true} along with the coefficients of a linear fit to the data. These fitted parameters allowed us to define an approximate transformation of the Einstein radius taking us from one class of source reconstruction to another. We found that the Einstein radius was the key parameter in controlling which class of source reconstruction was obtained. Fig.~\ref{fig:pred_true_q} shows that in both cases of erroneous source reconstructions, the axis ratio of the lens is most often under-estimated, but it does not follow an easily predictable pattern in the same way as the Einstein radius. Fig.~\ref{fig:pred_true_phi} shows that there is no apparent relationship between the inferred orientation of the mass profile and its true value when either the under-magnified or over-magnified solution is found.

\begin{table}
	\centering
	\caption{Summary of the approximate transformations of Einstein radius linking the inferred values in a given class of solution to the true value.}
	\label{tab:transform}
	\begin{tabular}{lcr} 
		\hline
		CNN Prediction & Transformation & Updated Prior \\
		\hline
		Under & $\theta = \frac{1}{0.46} (\hat \theta_U + 0.08)$  & $\theta_E \sim U[\theta \pm 0.25]$ \\
		Recon & $\theta = \theta_R$ & $\theta_E \sim U[\theta \pm 0.25]$ \\
		Over & $\theta = \frac{1}{2.11} (\hat \theta_O - 0.15)$ & $\theta_E \sim U[\theta \pm 0.5]$ \\
		\hline
	\end{tabular}
\end{table}

The relationship between the unphysical reconstructions and the correct solution allowed us to rapidly generate source reconstructions without the need for a full optimisation of the lens model. Fig.~\ref{fig:pred_true} shows how the predicted value of the Einstein radius relates to the true value in each of the three classes of source reconstruction we are considering here. The coefficients of a linear fit to the data allow us to construct an approximate transformation of the predicted Einstein radius to the true value for a given system. As expected, in the case of successful source reconstructions, the inferred value for the Einstein radius very closely matches the true value. The under-magnified solutions have inferred Einstein radii, $\hat \theta_{U}$ that can be approximated as $\hat \theta_{\mathrm{U}} \approx 0.46 \theta_{\mathrm{E}} - 0.08$, where $\theta_E$ is the true value for the system. Similarly, in the case of over-magnified solutions, the inferred Einstein radii $\hat \theta_{O}$ can be approximated as $\hat \theta_{\mathrm{0}} \approx 2.11 \theta_{\mathrm{E}} + 0.16$. Using these approximate transformations, along with the true parameters describing the lens, we identified the regions of parameter space where we expect each class of source reconstruction to occur. Keeping the position, axis ratio and orientation of the lens fixed to the truth, we varied the Einstein radius around its expected value and computed the linear inversion for each sample. The inversion achieving the highest evidence is considered to be the solution and we record the source reconstruction and residual image for our catalogue of training data.

\begin{figure*}
    \includegraphics[width=17cm]{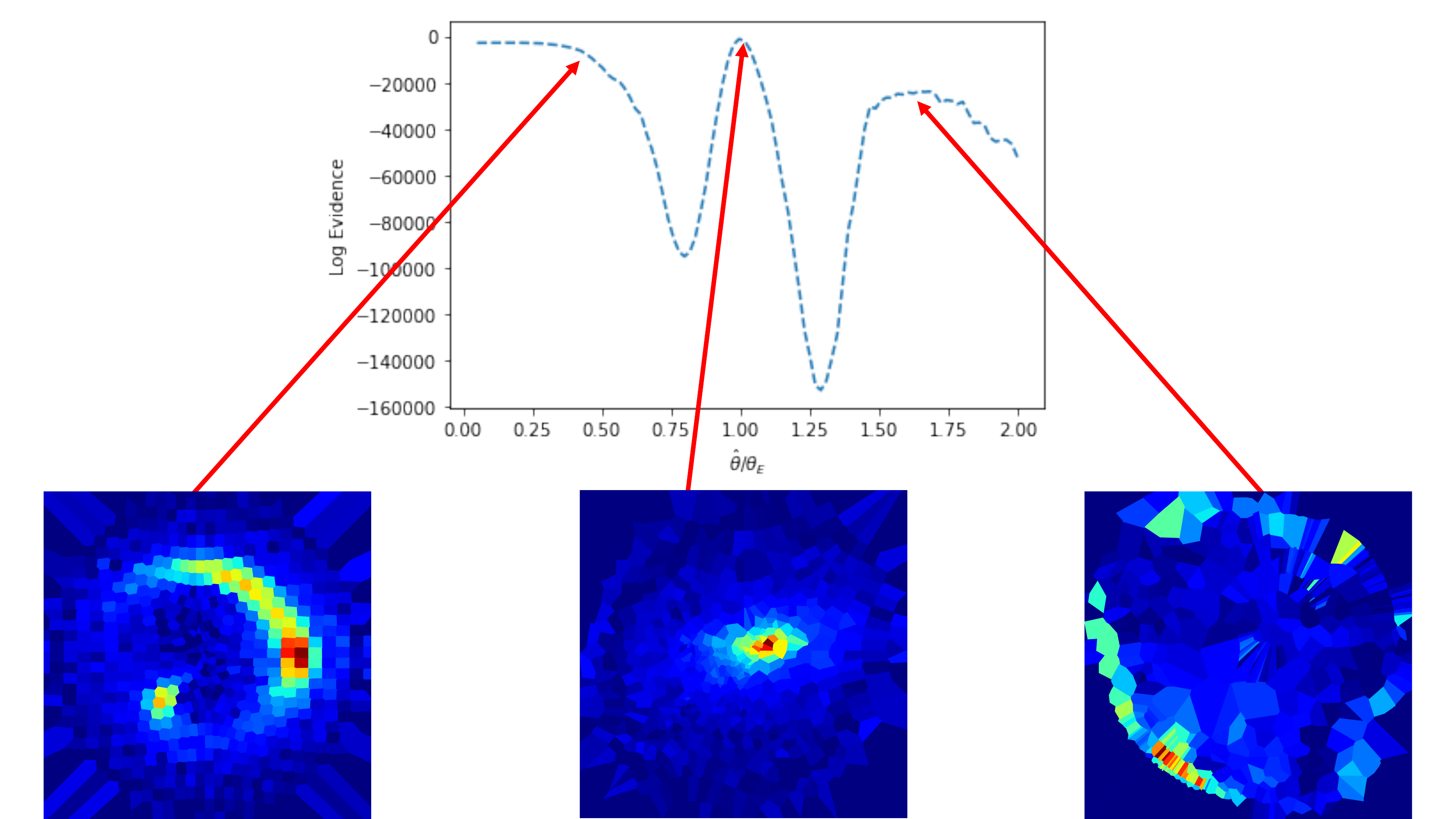}
    \caption{A one dimensional slice through Einstein radii values (normalised by the true Einstein radius), with all other mass model parameters fixed to their true values. Shown below from left to right: Under-magnified source reconstruction, successful source reconstruction, over-magnified source reconstruction.}
    \label{fig:1d}
\end{figure*}

\subsection{Testing Data}
\label{subsec:test}

\begin{figure}
    \includegraphics[width=\columnwidth]{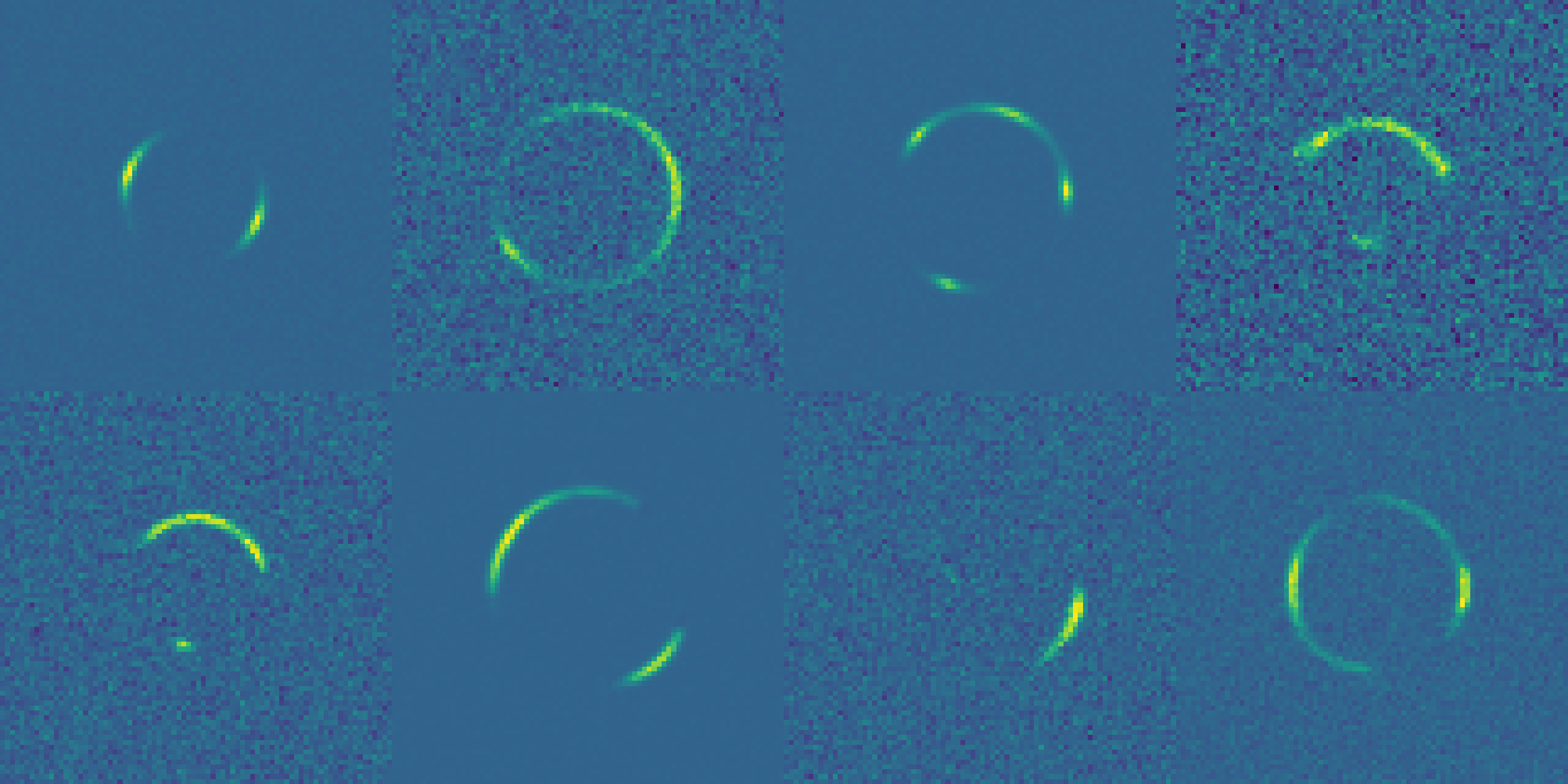}
    \caption{A selection of simulated SIE lenses with HUDF background sources, used to test the performance of the CNN on complex reconstructions. All images have a pixel scale of 0.1 arcsec pixel$^{-1}$ and each image's color scale has been normalised to the peak signal of the image.}
    \label{fig:pics}
\end{figure}

\begin{figure*}
    \includegraphics[width=17cm]{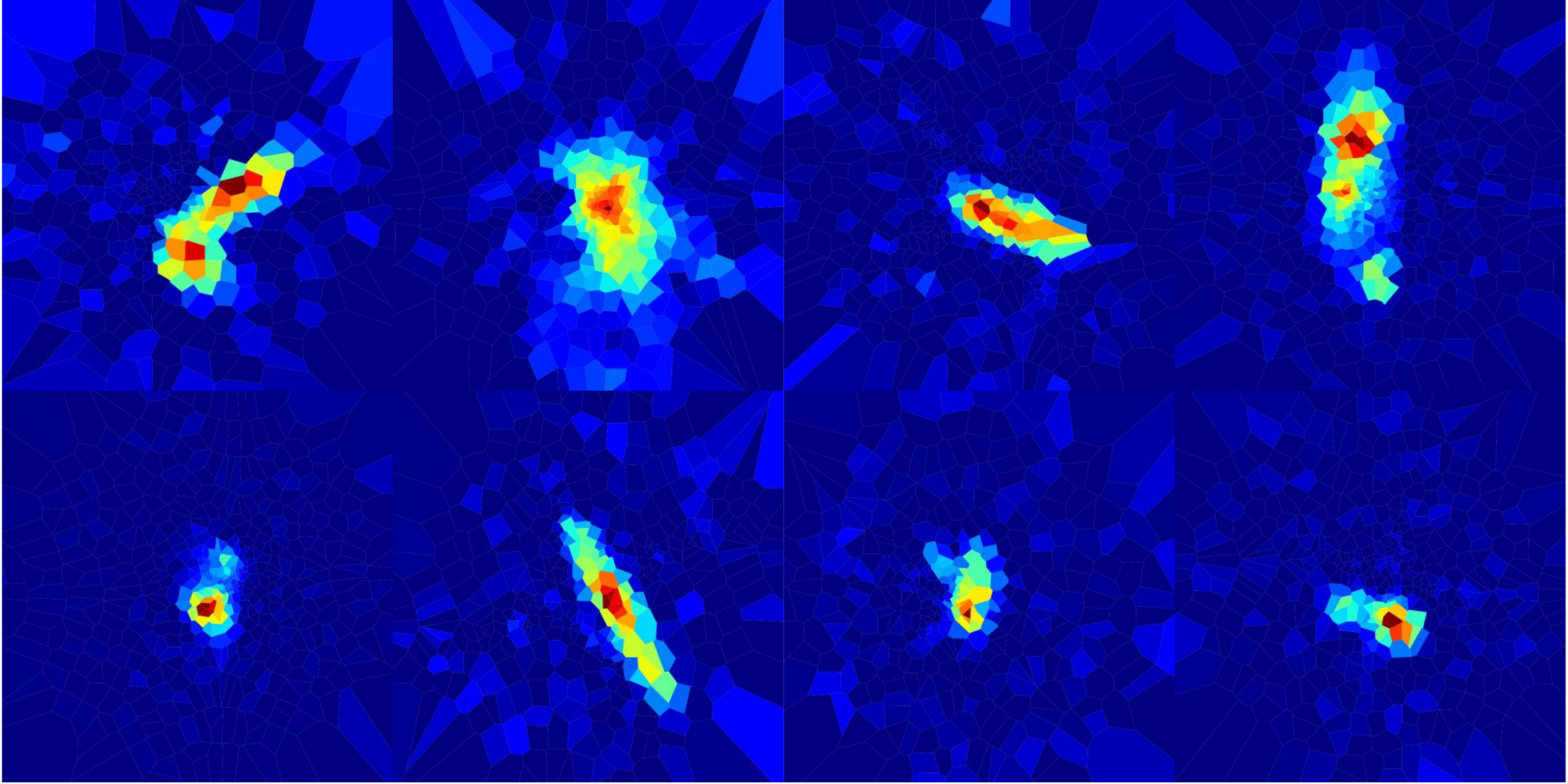}
    \caption{A selection of accurate source reconstructions of SIE-lensed HUDF galaxies. These reconstructions were used to test CNN performance on more complicated sources than the simple parametric sources used to create its training sample.}
    \label{fig:complex_recon}
\end{figure*}

A portion of the training data, produced as described in Section \ref{subsec:recon}, was set aside for evaluating the CNN's performance after training. These simple source reconstructions allowed us to test the network on a set of images with similar properties to the training data. In addition, to explore whether our CNN trained on reconstructions of simple parametric sources would be capable of classifying the reconstructions of more complex lensed sources, we produced SIE-lensed images of high redshift galaxies extracted from the Hubble Ultra Deep Field \citep[HUDF;][]{HUDF}. For this, we used the Pipeline for Images of Cosmological Strong lensing \citep[PICS;][]{pics}, simulating images to have the expected properties of Euclid VIS data \citep{cropper, niemi}. A sample of these simulated images is displayed in Fig.~\ref{fig:pics}. For each of these simulated images, we produced a source reconstruction corresponding to the under-magnified, over-magnified and accurate solution, following the same full analysis procedure described in Section \ref{subsec:recon}. These source reconstructions, along with the residual images of the models that produced them, were used to test the CNN's classification ability on significantly more complex images than it was trained on. A sample of the accurate HUDF source reconstructions is shown in Fig.~\ref{fig:complex_recon}.

\subsection{CNN Architecture}
\label{subsec:cnn}

Deep Neural Networks are a class of Artificial Neural Networks, consisting of multiple interconnected layers of nodes. The output of a node depends upon the weights of the connections made by the previous layer, as well as the bias of the current node. This information is fed into a non-linear activation function, controlling the strength of the output. CNNs are a further subset of neural networks built around multi-dimensional data. Convolutional filters, also known as kernels, are applied to the input to extract features from the data.

The network we built to classify our source reconstructions has a forked design, with two input paths. Each path consists of three convolutional layers and three max-pooling layers. The outputs of both paths are concatenated, before being flattened and fed into two fully connected layers. Dropout is employed between each layer to improve the network's resistance to over-fitting and the Leaky Rectified Linear Unit activation \citep[Leaky ReLU;][]{relu} function is used everywhere except for the final layer which employs the Sigmoid activation function. The Leaky ReLU activation function allows a small positive gradient for negative input values.

The tuneable hyper-parameters for our network, such as the number of convolutional layers, the size of the kernels and the dropout rates were set by a process of hyper-parameter optimisation. We opted to use Talos \citep{autonomio_2019} to automate the evaluation of model performance. In order to explore the very large parameter space, it was necessary to down-sample and look at a small fraction of combinations of parameter values. Once a rough estimate of hyper-parameters had been obtained, a more thorough search was carried out in a smaller region of parameter space. 

The network aims to predict the category of source reconstruction that a given input belongs to. To train the network, pairs of source reconstructions and the corresponding residual images are fed into the network in batches. The error on a prediction is determined via the categorical cross-entropy loss function (LOSS),
\begin{equation}
\mathrm{LOSS} = - \displaystyle \sum_{i=1}^{3} y_i \cdot \mathrm{log}\hat y_i
\end{equation}

where $y_i$ is the target value and $\hat y_i$ the predicted value. The network optimisation used the Nadam optimiser, which is a combination of stochastic gradient descent and Nesterov momentum \citep{nadam}. The CNN was trained and tested on a GPU machine, vastly improving the time taken to process large numbers of images. The training took place over 50 epochs, using 120,000 pairs of images.

\begin{figure*}
    \includegraphics[width=17cm]{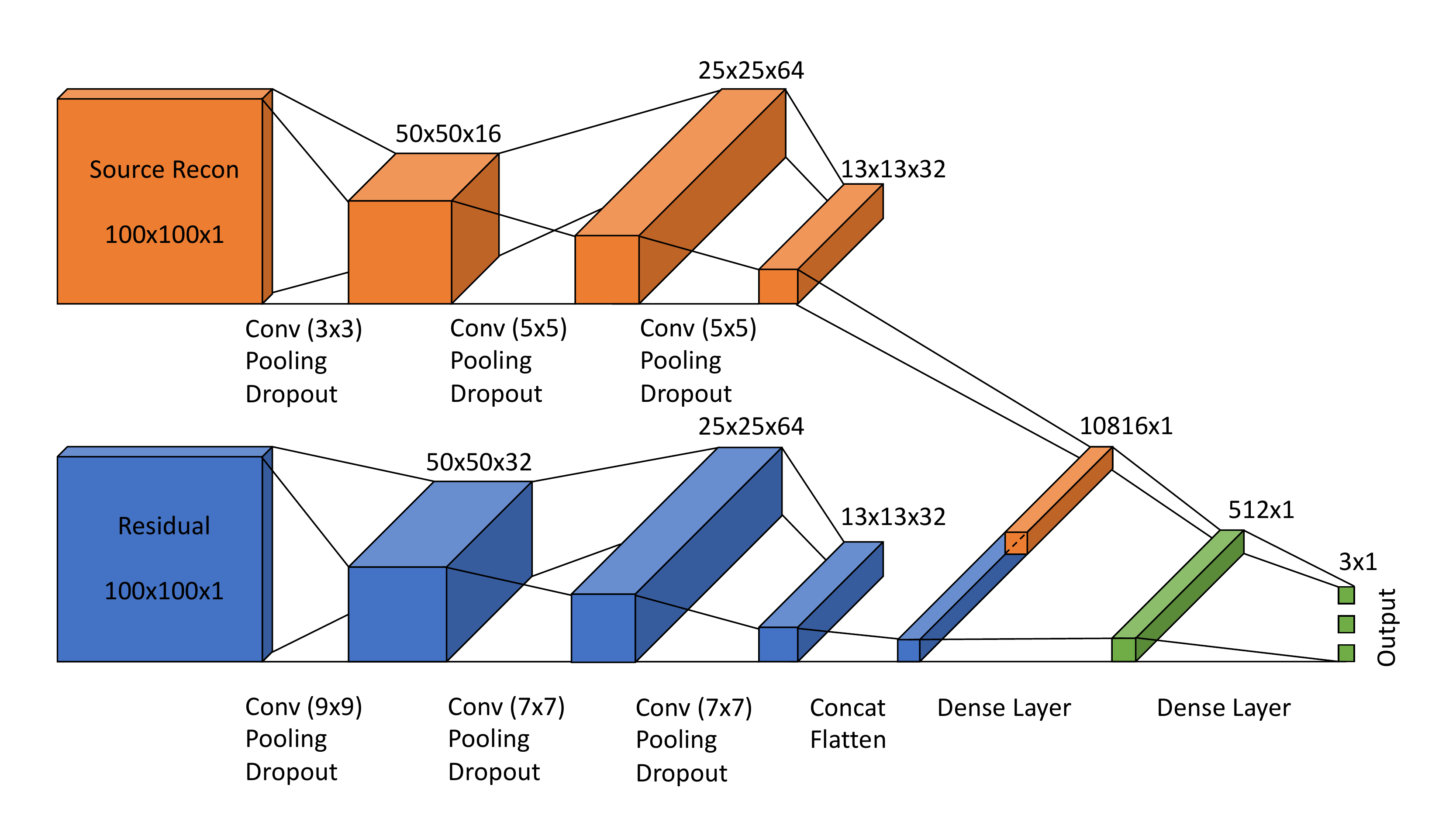}
    \caption{Structure of the CNN used in this work, showing the two input images and their respective paths in the network. There are six convolutional layers, each with max-pooling and dropout. A concatenation and flatten layer is included to join the outputs of the dual convolution pathways and connect this tensor with a 1D dense layer. LeakyReLU is used throughout the network, except for the activation of the final layer, which uses the Sigmoid activation function. The types of layers in the network at each step are given, along with the size of the kernel in pixels. The output dimensions are indicated above each block. A more detailed description can be found at the end of Section \ref{subsec:cnn}.}
    \label{fig:cnn_diagram}
\end{figure*}

The weights and biases of the network are summarised as follows:

\begin{itemize}
    \item Convolutional layer: For an input image of height $x_1$ and width $x_2$, the input is an ($x_1$, $x_2$, 1) matrix. The output of a convolutional layer is an ($x_1, x_2, N$) matrix, where N is the number of output filters applied in the convolution. Training adjusts the biases and weights for each filter, but their values remain fixed during each iteration. Each kernel of dimension ($k_1, k_2$) has an associated bias, giving a total of $k_1 \times k_2 \times N$ weights and N biases for each convolutional layer. The exact dimensions of each kernel are given in Fig.~\ref{fig:cnn_diagram}.
    
    \item Max-pooling layer: Pooling applies a 2 × 2 kernel with a stride of two, resulting in an output of dimension ($\left \lfloor{x_1/2}\right \rfloor$, $\left \lfloor{x_2/2}\right \rfloor$) for an input of ($x_1, x_2$).
    
    \item Concatenate: After the three convolutional layers in each input path of the network, the outputs are concatenated to form a tensor with dimensions (13, 13, 256). 
    
    \item First fully connected layer: The input is a flattened 43,264-node array, whilst the output is a 512-node array. Accordingly, there are $43,264 \times 512$ weights and 512 biases.
    
    \item Final layer: The input is an array with 512 nodes, whilst the output is a 3-node array (one node for each class of source reconstruction), hence there are 512 $\times$ 3 weights and 3 biases.
    
    \item There are a total of 5,820,323 trainable parameters.

\end{itemize}

\subsection{Combining CNN and Lens Modelling}
\label{subsec:combo}

The trained CNN is capable of taking a source reconstruction and a residual image, both of which are outputted in the lens modelling process, and returning an accurate prediction of whether the correct lens model has been found, or whether an under/over magnified solution has been identified. This prediction, along with the knowledge of how the inferred Einstein radius relates to each class of solution allows us to automatically correct the modelling process when erroneous solutions are found. Using the approximate transformations given in Table \ref{tab:transform} we can update the model's prior distribution on $\theta_E$ for subsequent modelling. In this way, we aim to improve the robustness of our modelling process against unwanted solutions and reduce the amount of human intervention required to produce accurate lens-models and source reconstructions. When considering the predictions of our CNN, we will use the abbreviation UM to refer to a predicted under-magnified solution, OM for a predicted over-magnified solution and C for when the network predicts a correct reconstruction.

To test this hybrid approach to lens modelling, we simulated a new set of 100 lensed images, following the approach detailed in Section \ref{subsec:sim}. We used {\tt PyAutoLens} to model each system, conducting a full analysis, allowing all the SIE mass-model parameters to vary and reconstructing the background source on a magnification based Voronoi grid. For all of the mass-model parameters, as well as the source plane pixelisation parameters, we opted to use uniform distributions covering a suitable range of parameter space. We chose a uniform prior distribution for the position of the lens centroid, centred on the true value with a width of 0.6 arcseconds. In the case of the orientation $\phi$ of the lens, we allowed the full range of values $\phi \sim U[0, \pi]$ radians. The axis ratio of the lens, $q$ was able to vary over the full range of values included in the simulations $q \sim U[0.25, 0.999]$. Again, the prior distribution of the Einstein radius $\theta_E$ followed a uniform distribution constrained only by the dimensions of the annular mask (computed according to the criteria detailed in \ref{subsec:recon}, $\theta_E \sim U[r_{min}, r_{max}]$. Such an approach to modelling the data was taken to show the extremes of how things can go wrong without some tuning of the priors before modelling begins. Furthermore, this serves to illustrate the problems experienced by sampling algorithms when exploring large and complex parameter spaces.

Once this initial round of modelling was completed, the source reconstruction and residual images were fed into our CNN to obtain a prediction on whether the modelling had been successful or not. The next step in the process depends on the prediction of the CNN as follows: 

\begin{itemize}
    \item UM prediction: The modelling process is repeated with an updated prior distribution on the Einstein radius. This new prior is defined in Table \ref{tab:transform}. The prior distributions on the other free parameters were left unchanged.
    
    \item C prediction: In this instance, we choose to repeat the modelling process with a decreased evidence tolerance and a narrowed uniform prior distribution centred on the inferred values from the previous modelling run. The goal of this repeated run is to more thoroughly explore the parameter space around the accepted solution and improve the accuracy of the model.
    
    \item OM prediction: The modelling process is repeated with an updated prior distribution on the Einstein radius, whilst leaving everything else unchanged. This new prior is defined in Table \ref{tab:transform}.

\end{itemize}

After this additional stage of modelling, the updated source reconstructions and residual images were fed into the CNN once more, providing a new prediction for each system. With this information, we proceeded similarly to before, but now we take into account the history of results for each system.

\begin{itemize}
    \item UM prediction:
    
    \begin{itemize}
        \item If the previous prediction was also UM, then the system is flagged for manual intervention at a later time. This indicates that the process for updating the priors was unable to move the model away from this solution, or that the CNN has misclassified a reconstruction.
        \item If the previous prediction was OM, this indicates that the prior update has 'overshot' the C solution and so a uniform prior on the Einstein radius is chosen to lie between the two previous values. The width of the prior was set such that it excludes the regions of parameter space that corresponded to the previous under and over-magnified solutions.
    \end{itemize}
    
    \item C prediction:
    
    \begin{itemize}
        \item If the previous prediction was UM, as before, we chose to repeat the modelling process with a decreased evidence tolerance and use narrowed uniform prior distributions centred on the inferred values from the previous modelling run. 
        \item if the previous prediction was C, no further action required.
        \item If the previous prediction was OM, again, we choose to repeat the modelling process with a decreased evidence tolerance and use narrowed uniform prior distributions centred on the inferred values from the previous modelling run. 
    \end{itemize}
    
    \item OM prediction: 
    
    \begin{itemize}
        \item If the previous prediction was also OM, then the system is flagged for manual intervention at a later time. This indicates that the process for updating the priors was unable to move the model away from this solution, or that the CNN has misclassified a reconstruction.
        \item If the previous prediction was UM, this indicates that the prior update 'overshot' the correct solution and so a uniform prior on the Einstein radius is chosen lying between the two previous values. The width of the prior is set such that it excludes the regions of parameter space that corresponded to the previous UM and OM solutions.
    \end{itemize}

\end{itemize}

This process can be repeated many times until an acceptable fraction of the CNN's predictions are that the correct model has been found. In practice, due to the crude nature of the prior-updating routine, there are diminishing returns on repeated cycles. The systems that become manually flagged during this process will need human intervention to guide the modelling to a suitable solution, but the overall load on the modeller is greatly reduced.

\section{Results}
\label{sec:results}

In this section, we present the results of testing our CNN on the reserved data-set, evaluating its performance on a per-class basis. We show that the CNN performs exceptionally well at the task of classifying source reconstructions. Additionally, we show the result of modelling 100 simulated observations using the procedure outlined in Section \ref{subsec:combo}. This set of images were simulated according to the procedures outlined in subsection \ref{subsec:sim}. Here, we opted to apply our iterative approach three times, observing good progress towards a complete sample of successfully modelled lenses with each step.

\subsection{CNN performance}

The CNN was trained on 130,000 pairs of source reconstructions and residual images, for 50 epochs. 10,000 pairs of source reconstructions and residual images were used as validation data throughout the training process. To further increase the variety in the training data, augmentation techniques were employed. Each pair of images was randomly reflected horizontally, vertically or rotated through an angle. The remaining 6,928 pairs of images were reserved as a testing set to evaluate the performance of the network on never before seen images once training had completed.

\begin{table}
	\centering
	\caption{Summary of key performance statistics for our CNN applied to testing data.}
	\label{tab:results}
	\begin{tabular}{lcccr} 
		\hline
		 & Precision & Recall & $F_1$-score & Support\\
		\hline
		UM & 0.9996 & 1.000 & 0.9998 & 2277\\
		C & 1.0000 & 0.9991 & 0.9996 & 2341\\
		OM & 0.9996 & 1.0000 & 0.9998 & 2310\\
		\hline
	\end{tabular}
\end{table}

\begin{figure}
	\includegraphics[width=\columnwidth]{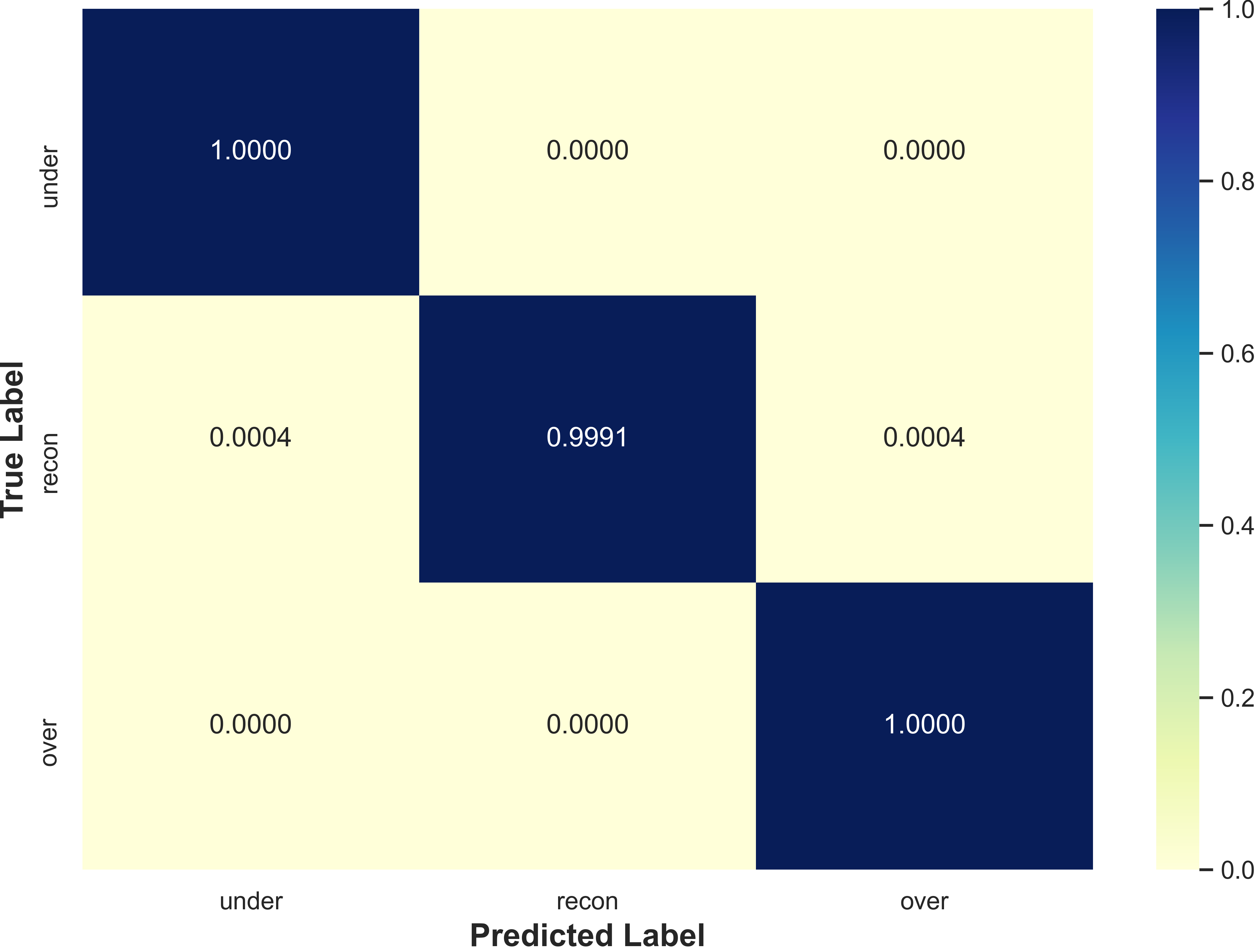}
    \caption{Confusion matrix for the CNN when tested on 6,928 never seen before pairs of source reconstructions and residual images for a simple Sérsic source. The confusion matrix has been normalised over its rows.}
    \label{fig:confusion}
\end{figure}

Fig.~\ref{fig:confusion} shows the confusion matrix $C$ for the CNN evaluated on the testing data set. The elements of this matrix are defined such that $C_{i,j}$ contains the number of true objects of class $i$ predicted to be in class $j$. Thus the diagonal elements of $C$ represent the correctly labelled instances and the off-diagonals where the network has incorrectly labelled an observation. The values displayed in $C$ are normalised over the rows. The CNN's recall or ability to find all samples of a particular class is above 99.9 per cent in all cases and performed perfectly on our test set for both under and over-magnified source reconstructions. Similarly, our CNN's precision, or ability to not label a sample of $Y$ as $X$ is greater than 99.9 per cent in all cases, with a perfect score in the case of successful source reconstructions. i.e, only successful source reconstructions were labelled as such. These results are summarised in Table \ref{tab:results}.

\begin{figure}
	\includegraphics[width=\columnwidth]{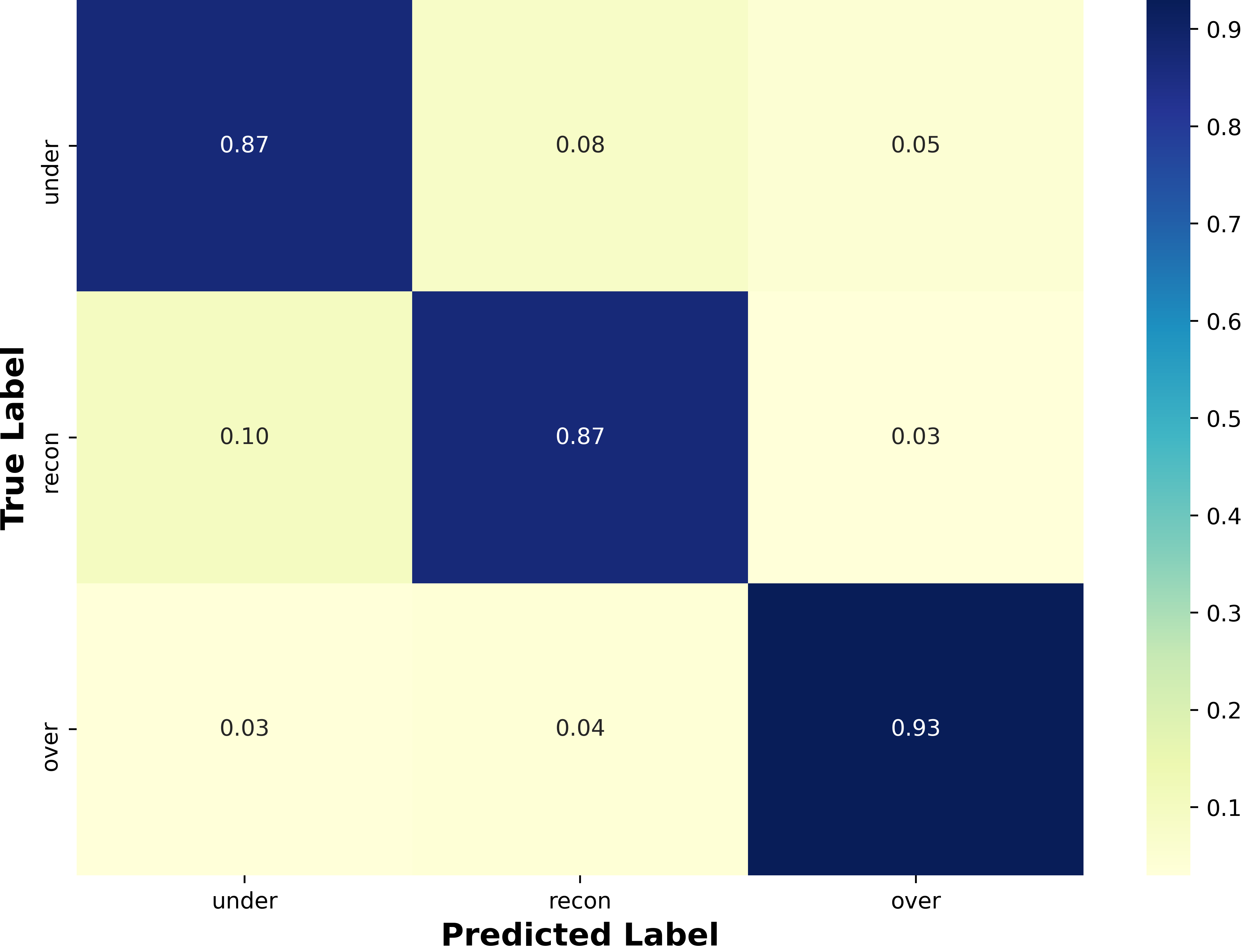}
    \caption{Confusion matrix for the CNN when trained on reconstructions of Sérsic sources and tested on reconstructions of HUDF sources. The confusion matrix has been normalised over its rows.
    }
    \label{fig:confusion_HUDF}
\end{figure}

As a further test of the CNN's ability to accurately classify source reconstructions, we applied it to the more complex HUDF source reconstructions described in Section \ref{subsec:test}. Here, the CNN gave predictions on 100 each of under-magnified, over-magnified and accurately reconstructed sources. We found that our CNN correctly classified 87 per cent of the under-magnified reconstructions, whilst misclassifying them as correctly reconstructed 8 per cent of the time, and incorrectly classifying 5 per cent of them as over-magnified. The CNN gave accurate predictions for 87 per cent of the correctly reconstructed sources, whilst incorrectly labelling 10 per cent as under-magnified and 3 per cent as over-magnified. Finally, the CNN correctly labelled 93 per cent of the over-magnified reconstructions, with just 3 per cent incorrectly labelled as under-magnified and 4 per cent mislabelled as accurate reconstructions. These results are summarised in Fig.~\ref{fig:confusion_HUDF}. The performance of the CNN on this complex dataset is remarkably good, given the simplicity of the reconstructed sources in the training data.

\subsection{Performance of {\tt PyAutoLens} combined with CNN}

Here, we describe the results of applying our CNN to blindly modelled data. For this, we have used our simulated images of Sérsic sources. We describe the process of using our CNN predictions to automatically adjust the prior distributions on the Einstein radius in three subsequent rounds of modelling. 

\begin{figure*}
    \includegraphics[width=17cm]{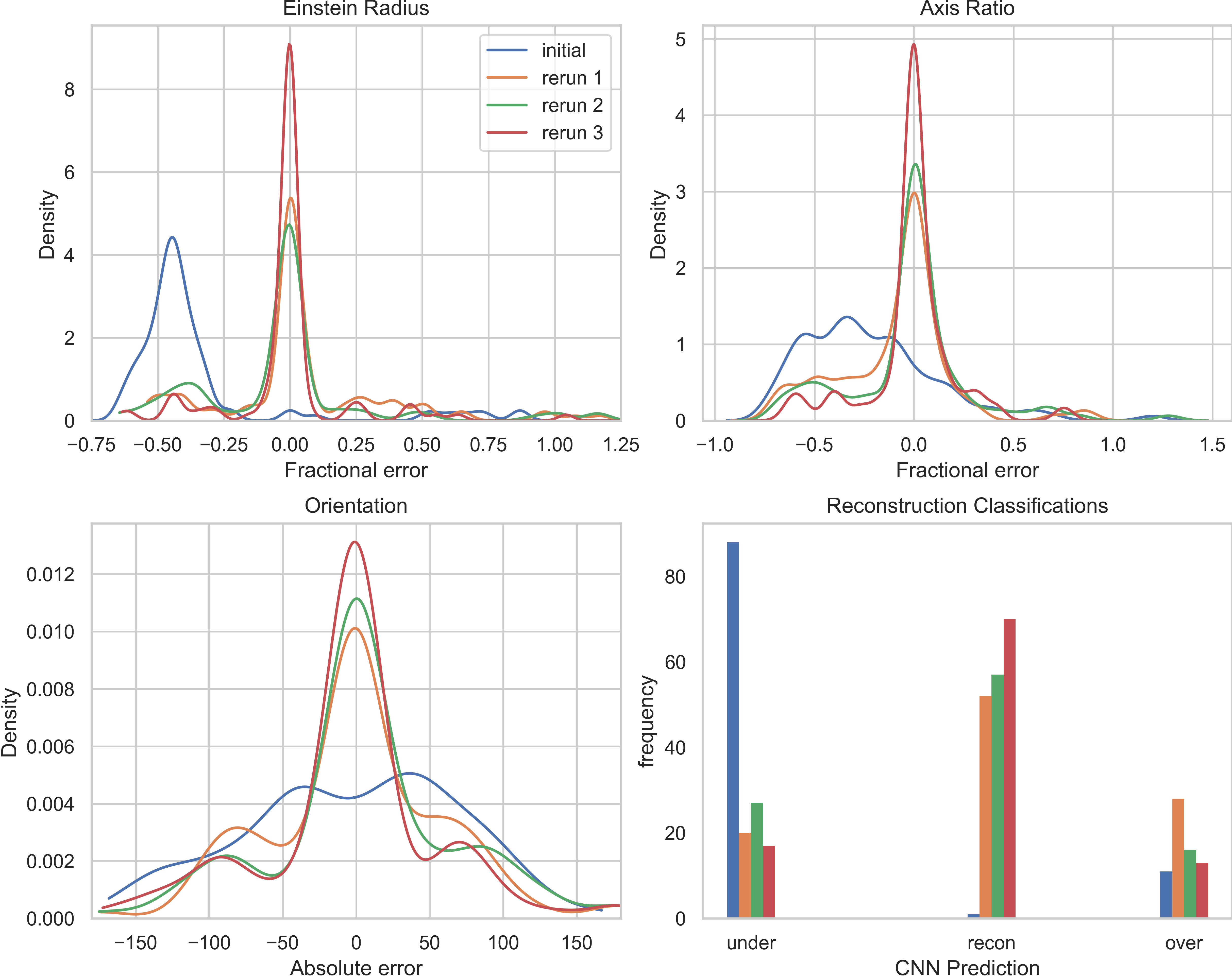}
    \caption{Top row: Fractional error distributions for the SIE mass model parameters, $\theta_E$ and $q$ after successive rounds of modelling. Bottom left: Absolute error distribution for the SIE mass model parameter $\phi$ after successive rounds of modelling. Bottom right: Histogram of the CNN's predictions for how the fraction of successful source reconstructions changes with successive rounds of modelling.}
    \label{fig:results_grid}
\end{figure*}

The results of this are presented in Fig.~\ref{fig:results_grid}. The initial modelling of this set of 100 lenses was carried out with no prior information on the lens model parameters and as such, under-magnified solutions have dominated the output. The bottom-right histogram in Fig.~\ref{fig:results_grid} shows how the proportion of different source reconstructions changes with each iteration of modelling according to our CNN predictions. Initially, our CNN identifies 88 models as UM, 11 as OM, and just 1 is identified as C. This is reflected in the error distributions for the key SIE mass model parameters. The top-left distribution in Fig.~\ref{fig:results_grid} shows the fractional error in Einstein radius $f_{\theta}$ for all 100 systems. There is a very significant peak in the initial data at $f_{\theta} = -0.45$, representing the large number of under-magnified solutions, and thus under-estimated Einstein radii. We also see in the top-right distribution of Fig.~\ref{fig:results_grid} the significant bias towards under-estimating the axis ratio of the lens. The bottom-left distribution, showing the absolute error on the inferred orientation of the lens, reflects the seemingly random relationship between the erroneous models and the true lens orientation. Labelled as rerun 1, rerun 2, and rerun 3, we show that the application of our CNN and prior updating routine to these results leads to a huge improvement in recovering the true lens parameters for the sample. After rerun 1 has been completed, much of the bias in the Einstein radii fractional error distributions is removed, though there is still significant density in regions indicating under and overestimation of its value. Similarly, in the case of the axis ratio, a clear peak around $f_q = 0$ has been formed, removing much of the probability mass in the under-estimate region of before. The inference of the orientation of the lens has also been greatly improved, as we would expect by increasing the number of successfully modelled systems. These results are reflected in the bottom right histogram of Fig.~\ref{fig:results_grid}, showing that the proportion of successful source reconstructions has increased from 1 to 52, according to our CNN predictions. The number of under-magnified reconstructions has been decreased by 68, down to just 20. The frequency of over-magnified solutions has increased, however, suggesting that our scheme for updating the Einstein radius prior has 'overshot' the correct solution in some cases. Rerun 2 increases the number of successful reconstructions by a small margin, but mostly results in moving solutions from the over-magnified category into the under-magnified category. Significant improvements are made in the final round of modelling, rerun 3, by considering the history of models for each case. For a system that has models that have previously been classified as under-magnified and over-magnified, we can search parameter space between the inferred Einstein radii values and hopefully converge upon the correct solution. In all of the error distributions for the mass-model parameters, we see improvements, i.e taller, narrower peaks centred on zero. After the final round of modelling, we achieved a decrease of 69 per cent in the occurrence overall of unphysical source reconstructions. The final count of successful source reconstructions stands at 70, with 17 under-magnified and 13 over-magnified solutions. In principle, we could continue with this process until we no longer see any improvement in the number of successful source reconstructions being identified by the CNN, or all systems that have not been labelled as C become flagged for manual inspection.

\section{Conclusions}
\label{sec:conclusions}

Strong gravitational lensing allows us to probe the mass distributions of the lensing galaxy as well as the properties of the background sources. Upcoming surveys such as LSST and Euclid are expected to observe in excess of one hundred thousand strong gravitational lenses. To deal with this huge amount of data, it is necessary to develop fast, robust and automatic lens modelling pipelines that do not require significant time investment from humans for each system. For this reason, we constructed a CNN to detect when the modelling process has gone awry and developed a simple scheme for automatically adjusting the prior distribution on the Einstein radius to guide the sampler to the correct solution. Simulated images with the resolution and expected seeing characteristics of the Euclid VIS instrument were created, to be used as inputs for the production of source reconstructions. We chose to simulate all of our lenses as SIEs and used Sérsic profiles for our sources. In both cases, we used realistic distributions of parameters that matched those observed in the SLACS survey. From these simulated images, we produced three source reconstructions for each observation corresponding to the under/over magnified solution and the correct solution. These source reconstructions, along with the residual images for the model were used to train a CNN to classify source reconstructions. We then blindly modelled 100 strong lenses, reconstructing the background sources on a Voronoi grid. The CNN was used to detect the kind of source reconstruction that had been produced, and this information coupled with a simple scheme for updating the prior distribution on the Einstein radius was used to improve upon the fraction of successfully modelled systems.

We find that our CNN is capable of extremely accurate identification of under-magnified, successful and over-magnified reconstructed sources. The network achieves a precision and recall over 99.9 per cent,  as well as an $f_1$-score, or harmonic mean of the precision and recall, greater than 0.99 across all classes of source reconstruction. In addition to identifying the class of solution that has been found, we have shown that a simple procedure for updating the model based on its predicted class can lead to significant improvements in the outcomes of blind modelling without the need for human intervention throughout the process. 

The success of our CNN in this task suggests that our procedure for generating the source reconstructions, omitting the full exploration of parameter space, has not negatively impacted its ability to perform the task. The axis ratio of the SIE mass-model corresponding to an erroneous solution tends to be under-estimated. Our network is trained on source reconstructions produced by fixing the axis ratio to its true value. This leads to the network being trained on images produced by less elliptical lens models than it might encounter when being tested upon a freely varied model.

It is possible that incorporating the information regarding erroneous source reconstructions tendency to have an under-estimated lens axis ratio could lead to improvements in our procedure for updating the model priors. An approach that uses a Gaussian prior to bias towards higher values of $q$, but with a standard deviation large enough to easily allow the exploration of the lower end of parameter space is something that could be investigated.

We have also tested our CNN, trained on reconstructions simple Sérsic sources, on reconstructions of images generated using real sources extracted from the HUDF. The CNN continued to perform well, showing that it can generalise to a more complex dataset without any retraining. There is however an obvious detriment to the performance of the network, and so the construction of a more complex training set would likely be beneficial. Before this technique can be applied to real data, further investigations into how our simplifications affect the network's performance are needed. One such simplification that we made was to omit lens light from our simulated images. Even in the best possible scenario of lens light removal, its presence will affect the noise characteristics of the image, which can impact the source reconstruction. Realistic features in our simulated images such as cosmic rays and hot pixels were not considered. Increased complexity of the sources in our training data would be required to deal with the variety of real images that might be observed and to minimise the performance decrease due to an overly simplified training set. Furthermore, a wider variety of mass models, the inclusion of substructure and deflectors along the line of sight would be required to create more realistic lensing scenarios. The question of how well this method of applying CNN predictions to parametric models generalises to real data requires further investigation.

\section*{Acknowledgements}

JM and NL acknowledge the support of the UK Science and Technology Facilities Council (STFC). SD is supported by a UK STFC Rutherford Fellowship.

\section*{Data Availability}

The data underlying this article will be shared on reasonable request to the corresponding author.



\bibliographystyle{mnras}
\bibliography{example} 

\begin{thebibliography}{}
\makeatletter
\relax
\def\mn@urlcharsother{\let\do\@makeother \do\$\do\&\do\#\do\^\do\_\do\%\do\~}
\def\mn@doi{\begingroup\mn@urlcharsother \@ifnextchar [ {\mn@doi@}
  {\mn@doi@[]}}
\def\mn@doi@[#1]#2{\def\@tempa{#1}\ifx\@tempa\@empty \href
  {http://dx.doi.org/#2} {doi:#2}\else \href {http://dx.doi.org/#2} {#1}\fi
  \endgroup}
\def\mn@eprint#1#2{\mn@eprint@#1:#2::\@nil}
\def\mn@eprint@arXiv#1{\href {http://arxiv.org/abs/#1} {{\tt arXiv:#1}}}
\def\mn@eprint@dblp#1{\href {http://dblp.uni-trier.de/rec/bibtex/#1.xml}
  {dblp:#1}}
\def\mn@eprint@#1:#2:#3:#4\@nil{\def\@tempa {#1}\def\@tempb {#2}\def\@tempc
  {#3}\ifx \@tempc \@empty \let \@tempc \@tempb \let \@tempb \@tempa \fi \ifx
  \@tempb \@empty \def\@tempb {arXiv}\fi \@ifundefined
  {mn@eprint@\@tempb}{\@tempb:\@tempc}{\expandafter \expandafter \csname
  mn@eprint@\@tempb\endcsname \expandafter{\@tempc}}}

\bibitem[\protect\citeauthoryear{Autonomio}{Autonomio}{2019}]{autonomio_2019}
Autonomio 2019, autonomio/talos, \url {http://www.github.com/autonomio/talos}

\bibitem[\protect\citeauthoryear{{Beckwith} et~al.,}{{Beckwith}
  et~al.}{2006}]{HUDF}
{Beckwith} S. V.~W.,  et~al., 2006, \mn@doi [\aj] {10.1086/507302}, \href
  {https://ui.adsabs.harvard.edu/abs/2006AJ....132.1729B} {132, 1729}

\bibitem[\protect\citeauthoryear{{Birrer}, {Amara}  \& {Refregier}}{{Birrer}
  et~al.}{2015}]{shapelets_birrer}
{Birrer} S.,  {Amara} A.,   {Refregier} A.,  2015, \mn@doi [\apj]
  {10.1088/0004-637X/813/2/102}, \href
  {https://ui.adsabs.harvard.edu/abs/2015ApJ...813..102B} {813, 102}

\bibitem[\protect\citeauthoryear{{Birrer} et~al.,}{{Birrer}
  et~al.}{2020}]{tdcosmo}
{Birrer} S.,  et~al., 2020, \mn@doi [\aap] {10.1051/0004-6361/202038861}, \href
  {https://ui.adsabs.harvard.edu/abs/2020A&A...643A.165B} {643, A165}

\bibitem[\protect\citeauthoryear{Bolton, Burles, Koopmans, Treu  \&
  Moustakas}{Bolton et~al.}{2006}]{bolton2006sloan}
Bolton A.~S.,  Burles S.,  Koopmans L.~V.,  Treu T.,   Moustakas L.~A.,  2006,
  The Astrophysical Journal, 638, 703

\bibitem[\protect\citeauthoryear{{Bolton}, {Burles}, {Koopmans}, {Treu},
  {Gavazzi}, {Moustakas}, {Wayth}  \& {Schlegel}}{{Bolton}
  et~al.}{2008}]{bolton2008sloan}
{Bolton} A.~S.,  {Burles} S.,  {Koopmans} L. V.~E.,  {Treu} T.,  {Gavazzi} R.,
  {Moustakas} L.~A.,  {Wayth} R.,   {Schlegel} D.~J.,  2008, \mn@doi [\apj]
  {10.1086/589327}, \href
  {https://ui.adsabs.harvard.edu/abs/2008ApJ...682..964B} {682, 964}

\bibitem[\protect\citeauthoryear{{Bom, C. R.}, {Makler, M.}, {Albuquerque, M.
  P.}  \& {Brandt, C. H.}}{{Bom, C. R.} et~al.}{2017}]{bom}
{Bom, C. R.} {Makler, M.} {Albuquerque, M. P.}  {Brandt, C. H.} 2017, \mn@doi
  [A\&A] {10.1051/0004-6361/201629159}, 597, A135

\bibitem[\protect\citeauthoryear{Brownstein et~al.,}{Brownstein
  et~al.}{2011}]{bells}
Brownstein J.~R.,  et~al., 2011, \mn@doi [The Astrophysical Journal]
  {10.1088/0004-637x/744/1/41}, 744, 41

\bibitem[\protect\citeauthoryear{{Cabanac, R. A.} et~al.,}{{Cabanac, R. A.}
  et~al.}{2007}]{CFHTLS}
{Cabanac, R. A.} et~al., 2007, \mn@doi [A\&A] {10.1051/0004-6361:20065810},
  461, 813

\bibitem[\protect\citeauthoryear{Cheng, Li, Conselice, Aragón-Salamanca, Dye
  \& Metcalf}{Cheng et~al.}{2020}]{sunny}
Cheng T.-Y.,  Li N.,  Conselice C.~J.,  Aragón-Salamanca A.,  Dye S.,
  Metcalf R.~B.,  2020, \mn@doi [Monthly Notices of the Royal Astronomical
  Society] {10.1093/mnras/staa1015}, 494, 3750

\bibitem[\protect\citeauthoryear{Collett}{Collett}{2015}]{collet}
Collett T.~E.,  2015, \mn@doi [The Astrophysical Journal]
  {10.1088/0004-637x/811/1/20}, 811, 20

\bibitem[\protect\citeauthoryear{Cropper et~al.,}{Cropper
  et~al.}{2016}]{cropper}
Cropper M.,  et~al., 2016, in MacEwen H.~A.,  Fazio G.~G.,  Lystrup M.,
  Batalha N.,  Siegler N.,   Tong E.~C.,  eds, ~ Vol. 9904, Space Telescopes
  and Instrumentation 2016: Optical, Infrared, and Millimeter Wave. SPIE, pp
  269 -- 284, \mn@doi{10.1117/12.2234739}, \url
  {https://doi.org/10.1117/12.2234739}

\bibitem[\protect\citeauthoryear{Dozat}{Dozat}{2015}]{nadam}
Dozat T.,  2015, In Proc. ICLR Workshop

\bibitem[\protect\citeauthoryear{Dye et~al.,}{Dye et~al.}{2015}]{sdp81}
Dye S.,  et~al., 2015, \mn@doi [Monthly Notices of the Royal Astronomical
  Society] {10.1093/mnras/stv1442}, 452, 2258

\bibitem[\protect\citeauthoryear{Dye et~al.,}{Dye et~al.}{2018}]{dye2018}
Dye S.,  et~al., 2018, \mn@doi [Monthly Notices of the Royal Astronomical
  Society] {10.1093/mnras/sty513}, 476, 4383

\bibitem[\protect\citeauthoryear{Feroz, Hobson  \& Bridges}{Feroz
  et~al.}{2009}]{multinest}
Feroz F.,  Hobson M.~P.,   Bridges M.,  2009, \mn@doi [Monthly Notices of the
  Royal Astronomical Society] {10.1111/j.1365-2966.2009.14548.x}, 398, 1601

\bibitem[\protect\citeauthoryear{Gavazzi, Marshall, Treu  \&
  Sonnenfeld}{Gavazzi et~al.}{2014}]{Gavazzi_2014}
Gavazzi R.,  Marshall P.~J.,  Treu T.,   Sonnenfeld A.,  2014, \mn@doi [The
  Astrophysical Journal] {10.1088/0004-637x/785/2/144}, 785, 144

\bibitem[\protect\citeauthoryear{{Hezaveh}, {Perreault Levasseur}  \&
  {Marshall}}{{Hezaveh} et~al.}{2017}]{hezaveh}
{Hezaveh} Y.~D.,  {Perreault Levasseur} L.,   {Marshall} P.~J.,  2017, \mn@doi
  [\nat] {10.1038/nature23463}, \href
  {https://ui.adsabs.harvard.edu/abs/2017Natur.548..555H} {548, 555}

\bibitem[\protect\citeauthoryear{{Ivezic} et~al.,}{{Ivezic}
  et~al.}{2008}]{lsst}
{Ivezic} Z.,  et~al., 2008, \mn@doi [Serbian Astronomical Journal]
  {10.2298/SAJ0876001I}, \href
  {https://ui.adsabs.harvard.edu/abs/2008SerAJ.176....1I} {176, 1}

\bibitem[\protect\citeauthoryear{{Jones}, {Swinbank}, {Ellis}, {Richard}  \&
  {Stark}}{{Jones} et~al.}{2010}]{jones_2010}
{Jones} T.~A.,  {Swinbank} A.~M.,  {Ellis} R.~S.,  {Richard} J.,   {Stark}
  D.~P.,  2010, \mn@doi [\mnras] {10.1111/j.1365-2966.2010.16378.x}, \href
  {https://ui.adsabs.harvard.edu/abs/2010MNRAS.404.1247J} {404, 1247}

\bibitem[\protect\citeauthoryear{Keeton}{Keeton}{2001}]{keeton2001catalog}
Keeton C.~R.,  2001, arXiv preprint astro-ph/0102341

\bibitem[\protect\citeauthoryear{Koopmans \& Treu}{Koopmans \&
  Treu}{2003}]{Koopmans_2003}
Koopmans L. V.~E.,  Treu T.,  2003, \mn@doi [The Astrophysical Journal]
  {10.1086/345423}, 583, 606

\bibitem[\protect\citeauthoryear{{Koopmans}, {Treu}, {Bolton}, {Burles}  \&
  {Moustakas}}{{Koopmans} et~al.}{2006}]{koopmans2006}
{Koopmans} L. V.~E.,  {Treu} T.,  {Bolton} A.~S.,  {Burles} S.,   {Moustakas}
  L.~A.,  2006, \mn@doi [\apj] {10.1086/505696}, \href
  {https://ui.adsabs.harvard.edu/abs/2006ApJ...649..599K} {649, 599}

\bibitem[\protect\citeauthoryear{{Lagattuta} et~al.,}{{Lagattuta}
  et~al.}{2010}]{lagattuta}
{Lagattuta} D.~J.,  et~al., 2010, \mn@doi [\apj]
  {10.1088/0004-637X/716/2/1579}, \href
  {https://ui.adsabs.harvard.edu/abs/2010ApJ...716.1579L} {716, 1579}

\bibitem[\protect\citeauthoryear{Laureijs et~al.}{Laureijs
  et~al.}{2011}]{euclid}
Laureijs R.,  et~al., 2011, arXiv

\bibitem[\protect\citeauthoryear{Levasseur, Hezaveh  \& Wechsler}{Levasseur
  et~al.}{2017}]{levasseur}
Levasseur L.~P.,  Hezaveh Y.~D.,   Wechsler R.~H.,  2017, \mn@doi [The
  Astrophysical Journal] {10.3847/2041-8213/aa9704}, 850, L7

\bibitem[\protect\citeauthoryear{Li \& Lee}{Li \& Lee}{1993}]{threshold}
Li C.,  Lee C.,  1993, \mn@doi [Pattern Recognition]
  {https://doi.org/10.1016/0031-3203(93)90115-D}, 26, 617

\bibitem[\protect\citeauthoryear{Li, Gladders, Rangel, Florian, Bleem,
  Heitmann, Habib  \& Fasel}{Li et~al.}{2016}]{pics}
Li N.,  Gladders M.~D.,  Rangel E.~M.,  Florian M.~K.,  Bleem L.~E.,  Heitmann
  K.,  Habib S.,   Fasel P.,  2016, \mn@doi [The Astrophysical Journal]
  {10.3847/0004-637x/828/1/54}, 828, 54

\bibitem[\protect\citeauthoryear{Marshall, Hogg, Moustakas, Fassnacht,
  Brada{\v{c}}, Schrabback  \& Blandford}{Marshall
  et~al.}{2009}]{Marshall_2009}
Marshall P.~J.,  Hogg D.~W.,  Moustakas L.~A.,  Fassnacht C.~D.,  Brada{\v{c}}
  M.,  Schrabback T.,   Blandford R.~D.,  2009, \mn@doi [The Astrophysical
  Journal] {10.1088/0004-637x/694/2/924}, 694, 924

\bibitem[\protect\citeauthoryear{{Morningstar} et~al.,}{{Morningstar}
  et~al.}{2019}]{morningstar}
{Morningstar} W.~R.,  et~al., 2019, \mn@doi [\apj] {10.3847/1538-4357/ab35d7},
  \href {https://ui.adsabs.harvard.edu/abs/2019ApJ...883...14M} {883, 14}

\bibitem[\protect\citeauthoryear{Nair \& Hinton}{Nair \& Hinton}{2010}]{relu}
Nair V.,  Hinton G.~E.,  2010, in Proceedings of the 27th International
  Conference on International Conference on Machine Learning. ICML'10.
Omnipress, Madison, WI, USA, p. 807–814

\bibitem[\protect\citeauthoryear{{Newton}, {Marshall}, {Treu}, {Auger},
  {Gavazzi}, {Bolton}, {Koopmans}  \& {Moustakas}}{{Newton}
  et~al.}{2011}]{newtonapj}
{Newton} E.~R.,  {Marshall} P.~J.,  {Treu} T.,  {Auger} M.~W.,  {Gavazzi} R.,
  {Bolton} A.~S.,  {Koopmans} L. V.~E.,   {Moustakas} L.~A.,  2011, \mn@doi
  [\apj] {10.1088/0004-637X/734/2/104}, \href
  {https://ui.adsabs.harvard.edu/abs/2011ApJ...734..104N} {734, 104}

\bibitem[\protect\citeauthoryear{Niemi}{Niemi}{2015}]{niemi}
Niemi S.-M.,  2015, Euclid Visible InStrument (VIS) Python Package (VIS-PP)
  Documentation, \url {https://www.mssl.ucl.ac.uk/~smn2/}

\bibitem[\protect\citeauthoryear{Nightingale \& Dye}{Nightingale \&
  Dye}{2015}]{adaptive_sli}
Nightingale J.~W.,  Dye S.,  2015, \mn@doi [Monthly Notices of the Royal
  Astronomical Society] {10.1093/mnras/stv1455}, 452, 2940

\bibitem[\protect\citeauthoryear{Nightingale \& Hayes}{Nightingale \&
  Hayes}{2020}]{pyautolens}
Nightingale J.,  Hayes R.,  2020, PyAutoLens: Open-source Strong Gravitational
  Lensing, \url {https://github.com/Jammy2211/PyAutoLens}

\bibitem[\protect\citeauthoryear{Nightingale, Dye  \& Massey}{Nightingale
  et~al.}{2018}]{Nightingale2018}
Nightingale J.~W.,  Dye S.,   Massey R.~J.,  2018, \mn@doi [Monthly Notices of
  the Royal Astronomical Society] {10.1093/mnras/sty1264}, 478, 4738

\bibitem[\protect\citeauthoryear{{Ostrovski} et~al.,}{{Ostrovski}
  et~al.}{2017}]{ostrovski}
{Ostrovski} F.,  et~al., 2017, \mn@doi [\mnras] {10.1093/mnras/stw2958}, \href
  {https://ui.adsabs.harvard.edu/abs/2017MNRAS.465.4325O} {465, 4325}

\bibitem[\protect\citeauthoryear{Park, Wagner-Carena, Birrer, Marshall, Lin  \&
  Roodman}{Park et~al.}{2020}]{Park2020}
Park J.~W.,  Wagner-Carena S.,  Birrer S.,  Marshall P.~J.,  Lin J. Y.-Y.,
  Roodman A.,  2020, arXiv

\bibitem[\protect\citeauthoryear{Pawase, Courbin, Faure, Kokotanekova  \&
  Meylan}{Pawase et~al.}{2014}]{pawase}
Pawase R.~S.,  Courbin F.,  Faure C.,  Kokotanekova R.,   Meylan G.,  2014,
  \mn@doi [Monthly Notices of the Royal Astronomical Society]
  {10.1093/mnras/stu179}, 439, 3392

\bibitem[\protect\citeauthoryear{Pearson, Li  \& Dye}{Pearson
  et~al.}{2019}]{pearson}
Pearson J.,  Li N.,   Dye S.,  2019, \mn@doi [Monthly Notices of the Royal
  Astronomical Society] {10.1093/mnras/stz1750}, 488, 991

\bibitem[\protect\citeauthoryear{{Planck Collaboration} et~al.,}{{Planck
  Collaboration} et~al.}{2016}]{planck2015}
{Planck Collaboration} et~al., 2016, \mn@doi [\aap]
  {10.1051/0004-6361/201525830}, \href
  {https://ui.adsabs.harvard.edu/abs/2016A&A...594A..13P} {594, A13}

\bibitem[\protect\citeauthoryear{{Richard}, {Jones}, {Ellis}, {Stark},
  {Livermore}  \& {Swinbank}}{{Richard} et~al.}{2011}]{richard_2011}
{Richard} J.,  {Jones} T.,  {Ellis} R.,  {Stark} D.~P.,  {Livermore} R.,
  {Swinbank} M.,  2011, \mn@doi [\mnras] {10.1111/j.1365-2966.2010.18161.x},
  \href {https://ui.adsabs.harvard.edu/abs/2011MNRAS.413..643R} {413, 643}

\bibitem[\protect\citeauthoryear{{Rizzo}, {Vegetti}, {Powell}, {Fraternali},
  {McKean}, {Stacey}  \& {White}}{{Rizzo} et~al.}{2020}]{rizzo}
{Rizzo} F.,  {Vegetti} S.,  {Powell} D.,  {Fraternali} F.,  {McKean} J.~P.,
  {Stacey} H.~R.,   {White} S.~D.~M.,  2020, \mn@doi [\nat]
  {10.1038/s41586-020-2572-6}, \href
  {https://ui.adsabs.harvard.edu/abs/2020Natur.584..201R} {584, 201}

\bibitem[\protect\citeauthoryear{{Seidel} \& {Bartelmann}}{{Seidel} \&
  {Bartelmann}}{2007}]{seidel}
{Seidel} G.,  {Bartelmann} M.,  2007, \mn@doi [\aap]
  {10.1051/0004-6361:20066097}, \href
  {https://ui.adsabs.harvard.edu/abs/2007A&A...472..341S} {472, 341}

\bibitem[\protect\citeauthoryear{Sonnenfeld, Treu, Gavazzi, Suyu, Marshall,
  Auger  \& Nipoti}{Sonnenfeld et~al.}{2013}]{evolution}
Sonnenfeld A.,  Treu T.,  Gavazzi R.,  Suyu S.~H.,  Marshall P.~J.,  Auger
  M.~W.,   Nipoti C.,  2013, \mn@doi [Astrophys. J.]
  {10.1088/0004-637X/777/2/98}, 777, 98

\bibitem[\protect\citeauthoryear{Sonnenfeld et~al.,}{Sonnenfeld
  et~al.}{2017}]{yattalens}
Sonnenfeld A.,  et~al., 2017, \mn@doi [Publications of the Astronomical Society
  of Japan] {10.1093/pasj/psx062}, 70

\bibitem[\protect\citeauthoryear{Suyu, Marshall, Hobson  \& Blandford}{Suyu
  et~al.}{2006}]{suyu}
Suyu S.~H.,  Marshall P.,  Hobson M.,   Blandford R.,  2006, \mn@doi [Mon. Not.
  Roy. Astron. Soc.] {10.1111/j.1365-2966.2006.10733.x}, 371, 983

\bibitem[\protect\citeauthoryear{{Swinbank} et~al.,}{{Swinbank}
  et~al.}{2009}]{swinbank_2009}
{Swinbank} A.~M.,  et~al., 2009, \mn@doi [\mnras]
  {10.1111/j.1365-2966.2009.15617.x}, \href
  {https://ui.adsabs.harvard.edu/abs/2009MNRAS.400.1121S} {400, 1121}

\bibitem[\protect\citeauthoryear{{Sygnet, J. F.}, {Tu, H.}, {Fort, B.}  \&
  {Gavazzi, R.}}{{Sygnet, J. F.} et~al.}{2010}]{sygnet}
{Sygnet, J. F.} {Tu, H.} {Fort, B.}  {Gavazzi, R.} 2010, \mn@doi [A\&A]
  {10.1051/0004-6361/200913977}, 517, A25

\bibitem[\protect\citeauthoryear{Tagore \& Jackson}{Tagore \&
  Jackson}{2016}]{shapelet_bayes}
Tagore A.~S.,  Jackson N.,  2016, \mn@doi [Monthly Notices of the Royal
  Astronomical Society] {10.1093/mnras/stw057}, 457, 3066

\bibitem[\protect\citeauthoryear{Vegetti \& Koopmans}{Vegetti \&
  Koopmans}{2009a}]{vegetti_koopmans_2009a}
Vegetti S.,  Koopmans L. V.~E.,  2009a, \mn@doi [Monthly Notices of the Royal
  Astronomical Society] {10.1111/j.1365-2966.2008.14005.x}, 392, 945

\bibitem[\protect\citeauthoryear{Vegetti \& Koopmans}{Vegetti \&
  Koopmans}{2009b}]{vegetti_koopmans_2009b}
Vegetti S.,  Koopmans L. V.~E.,  2009b, \mn@doi [Monthly Notices of the Royal
  Astronomical Society] {10.1111/j.1365-2966.2009.15559.x}, 400, 1583

\bibitem[\protect\citeauthoryear{{Vegetti}, {Despali}, {Lovell}  \&
  {Enzi}}{{Vegetti} et~al.}{2018}]{darkmatter}
{Vegetti} S.,  {Despali} G.,  {Lovell} M.~R.,   {Enzi} W.,  2018, \mn@doi
  [\mnras] {10.1093/mnras/sty2393}, \href
  {https://ui.adsabs.harvard.edu/abs/2018MNRAS.481.3661V} {481, 3661}

\bibitem[\protect\citeauthoryear{{Warren} \& {Dye}}{{Warren} \&
  {Dye}}{2003}]{SLI}
{Warren} S.~J.,  {Dye} S.,  2003, \mn@doi [\apj] {10.1086/375132}, \href
  {https://ui.adsabs.harvard.edu/abs/2003ApJ...590..673W} {590, 673}

\bibitem[\protect\citeauthoryear{Wong et~al.}{Wong et~al.}{2020}]{wong}
Wong K.~C.,  et~al., 2020, \mn@doi [Mon. Not. Roy. Astron. Soc.]
  {10.1093/mnras/stz3094}, 498, 1420

\makeatother
\end{thebibliography}





\bsp	
\label{lastpage}
\end{document}